\documentclass[prl,a4paper,showpacs,twocolumn,superscriptaddress,longbibliography]{revtex4-1}

\usepackage{amssymb}
\usepackage{amsmath}
\usepackage{amsfonts}
\usepackage{graphicx}
\usepackage{bm}
\usepackage{color}
\usepackage{multirow}
\usepackage{natbib}
\usepackage{hyperref}
\usepackage[normalem]{ulem}
\usepackage{relsize}
\usepackage[final]{pdfpages}
\usepackage{pgffor}

\DeclareMathOperator{\Tr}{Tr}

\DeclareMathOperator{\im}{Im}
\DeclareMathOperator{\re}{Re}

\makeatletter
\AtBeginDocument{\let\LS@rot\@undefined}
\makeatother

\begin{document}

\title{Dissipative and Hall viscosity of a disordered 2D electron gas}

\author{Igor S.~Burmistrov}

\affiliation{\hbox{L.~D.~Landau Institute for Theoretical Physics, acad. Semenova av. 1-a, 142432 Chernogolovka, Russia}}

\affiliation{Laboratory for Condensed Matter Physics, National Research University Higher School of Economics, 101000 Moscow, Russia}



\author{Moshe Goldstein}

\affiliation{Raymond and Beverly Sackler School of Physics and Astronomy, Tel Aviv University, Tel Aviv 6997801, Israel}

\author{Mordecai Kot}

\affiliation{Raymond and Beverly Sackler School of Physics and Astronomy, Tel Aviv University, Tel Aviv 6997801, Israel}

\author{Vladislav D.~Kurilovich}

\affiliation{Departments of Physics and Applied Physics, Yale University, New Haven, CT 06520, USA}


\author{Pavel D.~Kurilovich}

\affiliation{Departments of Physics and Applied Physics, Yale University, New Haven, CT 06520, USA}



\begin{abstract}
Hydrodynamic charge transport is at the center of recent research efforts. Of particular interest is the nondissipative Hall viscosity, which conveys topological information in clean gapped systems. The prevalence of disorder in the real world calls for a study of its effect on viscosity. Here we address this question, both analytically and numerically, in the context of a disordered noninteracting 2D electrons.
Analytically, we employ the self-consistent Born approximation,  explicitly taking into account the modification of the single-particle density of states and the elastic transport time due to the Landau quantization. The reported results interpolate smoothly between the limiting cases of weak (strong) magnetic field and strong (weak) disorder. In the regime of weak magnetic field our results describes the quantum (Shubnikov-de Haas type) oscillations of the dissipative and Hall viscosity. For strong magnetic fields we characterize the effects of the disorder-induced broadening of the Landau levels on the viscosity coefficients. This is supplemented by numerical calculations for a few filled Landau levels. Our results show that the Hall viscosity is surprisingly robust to disorder.
\end{abstract}

\maketitle

\textsf{Introduction.}\ --- Ordinary fluid motion is described by the theory of hydrodynamics, one of whose cornerstones is viscosity, which serves as the source of dissipation. Under certain conditions, charge transport in an electronic system can also be dominated by hydrodynamic viscous 
flow~\cite{Gurzhi,Molenkamp}. The discovery of graphene stimulated renewed theoretical~\cite{Muller2009,Andreev2011,Polini2015,Levitov2016,NGMS,Levitov2017,Kashuba,Lucas2018,Schmalian} and experimental~\cite{Titov2013,Polini2016,Kim2016,Moll2016,Geim2017,Levitov2018,Kvon2018} interest in the hydrodynamic description of charge conduction. 
 
In the absence of time-reversal symmetry the viscosity tensor has non-dissipative antisymmetric components. In the presence of a magnetic field $B$, this non-dissipative Hall viscosity ($\eta_H$) was studied theoretically in the classical limit of high temperature plasmas \cite{CC,Marshall,Kaufman,Thompson,Braginskii}, and for low temperature electron gas \cite{Steinberg}. Later,  interest in the Hall viscosity was rekindled in quantum systems with a gapped spectrum, due to the connection between $\eta_H$ and geometric response~\cite{Avron,Levay,Avron1998,Read,RR,Haldane,HLF}, and its expected quantization in the presence of translational and rotational symmetries~\cite{RR}. It was understood that beyond the Hall conductivity and viscosity there are additional non-dissipative electro-magnetic and geometrical response functions in gapped quantum systems~\cite{HS2012,Abanov2013,AG2014,GA2014,Hoyos,Andreev,Wiegmann2014,GA2015,Wiegmann2015,Gurarie,Andreev2015,NG2017,NCG2017}. Within the hydrodynamic description of electron transport, non-zero $\eta_H$ influences significantly the structure of the electron flow \cite{Alekseev,PTP,SNSMM,DG2017,GA2017,Alekseev-2}, which allows one to access $\eta_H$ experimentally \cite{Bandurin2018}. Also, it was argued that the dissipative and Hall viscosity affect the spectrum of edge magnetoplasmons~\cite{ACG2018,SDVV,CG}. 

For noninteracting electrons in the absence of disorder each filled Landau level (LL) gives a contribution to 
the Hall viscosity equal $\hbar (2n+1)/(8\pi l_B^2)$ \cite{Avron}, where $n$ denotes the LL index and $l_B= \sqrt{\hbar c/(e B)}$ stands for the magnetic length. This result is stable against perturbations of the Hamiltonian which preserve translational and rotational invariance \cite{RR}. However, the fate of this result in the presence of disorder has not been studied yet. Therefore, it is not clear how the clean result obtained within the quantum treatment of the electron motion in a magnetic field connects to the result $\eta_H = \nu_0 \mu^2 \omega_c\tau_{{\rm tr},2}^2/[1+4\omega_c^2\tau_{{\rm tr},2}^2]$ derived for a classical disordered electron gas~\cite{Steinberg}. Here $\mu$ denotes the chemical potential, $\tau_{{\rm tr},2}$ the second transport time, $\omega_c=eB/(m c)$ the cyclotron frequency, $\nu_0 = m/(2\pi\hbar^2)$ the density of states at $B=0$, and $m$ the effective electron mass. 

In this Letter we report the results of an analytical and numerical study of \emph{the dissipative and Hall viscosities} of noninteracting 2D electrons in the presence of \emph{disorder}. Contrary to previous studies we \emph{explicitly} take into account the \emph{Landau quantization} of the electron spectrum. Analytically, within the self-consistent Born approximation (SCBA) \cite{AFS} we derive expressions for the dissipative and Hall viscosities, which smoothly interpolates between the results known in the literature for classical magnetic field \cite{CC,Marshall,Kaufman,Thompson,Braginskii,Alekseev} and for the strong magnetic field in the absence of disorder \cite{Avron}. Since the SCBA is rigorously justified for high LLs only, we perform numerical calculation of $\eta_H$ for a few lowest LLs. The obtained numerical results are in a perfect agreement with the SCBA predictions. They demonstrate a surprising resilience of the Hall viscosity 
to disorder.

\textsf{Model.}\ --- Noninteracting electrons confined to a 2D plane are described by the following 
Hamiltonian
\begin{equation}
H = \bigl (- i\nabla - e \bm{A} \bigr )^2/2m + V(\bm{r}) , \label{eq:model}
\end{equation}
where 
$V(\bm{r})$ stands for a random potential and $\bm{A}$ for the vector potential corresponding to the static perpendicular magnetic field $B$. In this paper we use the Landau gauge: $A_y = - B x$ and $A_x=A_z=0$. We assume that the random potential has Gaussian distribution with a pair correlation function $\overline{V(\bm{r}) V(\bm{r^\prime})}= W(|\bm{r}-\bm{r^\prime}|)$ which decays with a typical length scale $d$.
In what follows we use units with $e=c=\hbar=k_B=1$.

\textsf{Kubo formula for the viscosity.}\ --- The viscosity tensor can be computed by means of the Kubo formula \cite{Resibois,McLennan,Bradlyn2012}:
\begin{align}
\eta_{jk,ps}(\omega) = & - \frac{i \kappa^{-1}}{\omega}\delta_{jk}\delta_{ps}+ \frac{1}{i \omega S}
\int \frac{d\Omega d\varepsilon}{\pi^2} 
\frac{ f_\varepsilon-f_{\varepsilon+\Omega}}
{\Omega-\omega-i0} 
 \notag \\
& \times
\overline{\Tr T_{jk} \im G^R_{\varepsilon+\Omega} T_{ps}
\im G^R_\varepsilon } .
\label{eq:Kubo}
\end{align}
Here $f_\varepsilon = 1/[1+\exp((\varepsilon-\mu)/T)]$ denotes the Fermi distribution function, $G^{R}_\varepsilon = 1/(\varepsilon-H-i0)$ the retarded Green's function,  $S$ the system area, and  $\kappa$ the internal compressibility \cite{Footnote1}.   The form of the stress tensor, 
$T_{jk}=m(v_jv_k+v_kv_j)/2$ is not affected by a random potential (see Ref. \cite{Irving1950} and the Supplemental Material for details~\cite{SM}).
Here $\bm{v} = (- i\nabla - e \bm{A})/m$ stands for the velocity operator.
\color{black}
Disorder averaging is denoted by an overbar. 

\textsf{Self-consistent Born approximation.}\ --- In order to compute the viscosity tensor from Eq.~\eqref{eq:Kubo} we treat the disorder scattering using the SCBA~\cite{AFS}. This approximation holds under the following conditions~\cite{RS,LA}:
\begin{equation}
1/k_F, d \ll l_B,\qquad d\ll v_F \tau_0 .
\label{eq:assumptions}
\end{equation}
 Here  $k_F=mv_F$ and $v_F$ denote the Fermi momentum and velocity, respectively, and $\tau_0$ is the total  elastic relaxation time at zero magnetic field. It can be expressed in terms of the Fourier transform $\tilde{W}(\bm{q})$ of the pair correlation function $W(\bm{r})$. Furthermore, it is convenient to generalize it to ($m=0,1,2,\dots$):
\begin{equation}
\frac{1}{\tau_m} = \nu_0 \int_0^{2\pi}d\phi\, \tilde{W}(2k_F \sin\phi/2) \cos m \phi .
\end{equation} 

The average density of states $\nu(\varepsilon)$ at non-zero $B$ is determined by the average retarded Green's function $\mathcal{G}^R_\varepsilon$. In the LL representation the average density of states is given as  
$\nu(\varepsilon) = -\sum_n \im \mathcal{G}^R_n(\varepsilon)/(2\pi^2 l_B^2)$. Within SCBA the retarded Green's function satisfies \cite{AFS,RS,LA}    
\begin{equation}
\mathcal{G}^R_n=(\varepsilon-\epsilon_n-\Sigma^R_\varepsilon)^{-1}, \quad 
\Sigma^R_\varepsilon = \frac{\omega_c}{2\pi \tau_0} 
\sum_n \mathcal{G}^R_n ,
\label{eq:self-consistent}
\end{equation}
where $\epsilon_n = \omega_c(n+1/2)$.
There are two limiting cases in which the self-consistent Eq.~\eqref{eq:self-consistent} can be easily solved~\cite{AFS}. In the regime of overlapping LLs, $\omega_c\tau_0\ll 1$, one can use the Poisson formula for summation over LL index. The averaged density of states becomes $\nu(\varepsilon) = \nu_0 [1- 2\delta \cos(2\pi \varepsilon/\omega_c)]$. Here 
$\delta = \exp(-\pi/\omega_c\tau_0) \ll 1$ is the Dingle parameter. In the opposite case, when the LLs are well separated, one can restrict the summation in Eq.~\eqref{eq:self-consistent} to the single LL which is closest to the energy of interest, $|\varepsilon - \epsilon_N|<\omega_c/2$. Then the average density of states acquires the semi-circle profile: $\nu(\varepsilon) = \nu_0 \tau_0 \sqrt{\Gamma^2-(\varepsilon-\epsilon_N)^2}$, where $\Gamma = \sqrt{2\omega_c/(\pi \tau_0)}$ determines the the broadened LL width.

\begin{figure}[t]
\centerline{\includegraphics[width=0.36\textwidth]{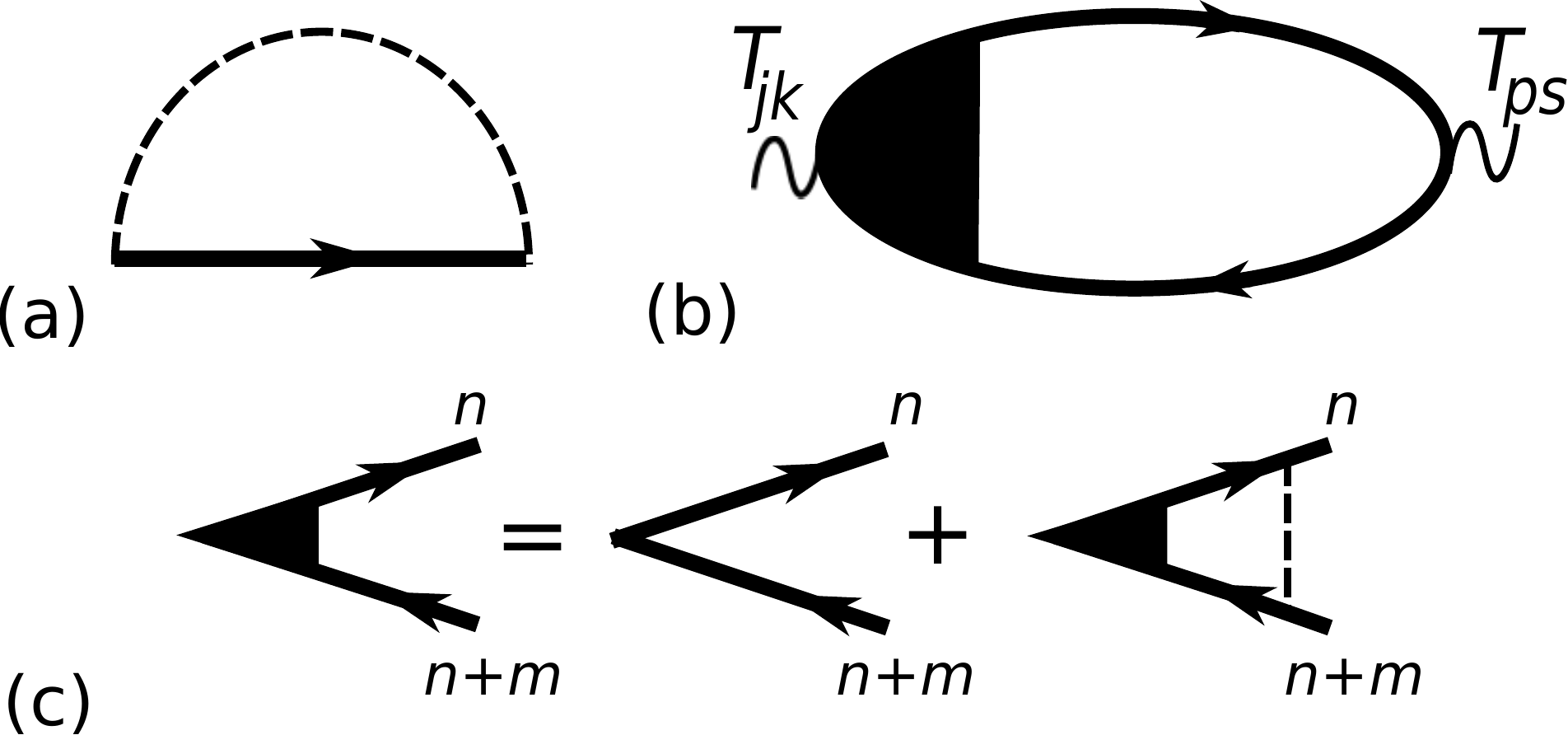}}
\caption{(a) The self-energy diagram. (b) The diagram corresponding to the Kubo formula~\eqref{eq:Kubo}. (c) The equation for the vertex $V_m$ in the ladder approximation.
The solid line denotes the SCBA Green's function $\mathcal{G}(\varepsilon)$. The dashed line denotes the disorder correlation function $W(\bm{r})$.
}
\label{Figure-0}
\end{figure}

In the presence of long-range disorder correlations, $d\gg k_F^{-1}$, it is important to take into account the vertex corrections to the ``bubble'' contribution in the Kubo formula~\eqref{eq:Kubo} (see Fig.~\ref{Figure-0}). This implies that in addition to the average Green's function, one also needs to know the renormalized vertex, which is the stress tensor in the case of the viscosity. Within the SCBA $T_{jk}$ can be approximated as a linear combination of operators which change the LL index by $2$. Under conditions~\eqref{eq:assumptions} one can show that an operator $V_m$, which transfers an electron from the $(n+m)$-th LL  to the $n$-th LL,  is renormalized by the ladder resummation of the disorder lines as follows \cite{RS,LA,DMP} (see~\cite{SM} for details):
\begin{equation}
V_m \to 
\frac{V_m}{1-\tau_m^{-1}\Pi_m^{RA}}, \,
\Pi_m^{RA} = \frac{\omega_c}{2\pi}
\sum_n \mathcal{G}^R_{n+m}(\varepsilon) 
\mathcal{G}^A_{n}(\varepsilon) .
\label{eq:vertex}
\end{equation}   
Here $\Pi_m^{RA}$ is the contribution of the bubble without ladder insertions. Using Eq.~\eqref{eq:self-consistent}, it can be rewritten as $\Pi_m^{RA}= \color{black} - \color{black} i \nu(\varepsilon)/[m \omega_c \nu_0 - i \nu(\varepsilon)\color{black}/\tau_0\color{black}]$. Therefore, within the SCBA the vertex corrections are expressed in terms of the average density of states only.

\textsf{Dissipative viscosity.}\ --- Disorder averaging restores 2D rotational symmetry~\cite{note:rotation}. Hence, 
the viscosity tensor $\eta_{jk,ps}$ is characterized by only three parameters: 
\begin{align}
\eta_{jk,ps}  & =   \eta_{s} \bigl (\delta_{jp}\delta_{ks}+
\delta_{js}\delta_{kp}\bigr ) + \bigl (\zeta- \eta_{s}\bigr )\delta_{jk}\delta_{ps} \notag \\
& + (\eta_H/2) \bigl (
\epsilon_{jp}\delta_{ks}+\epsilon_{js}\delta_{kp}+\epsilon_{kp}\delta_{js}+\epsilon_{ks}\delta_{jp}\bigr ) ,
\end{align}
where $\zeta$ and $\eta_{s}$ denotes the bulk and shear viscosities, respectively. Within the SCBA 
the bulk viscosity vanishes, $\zeta=0$. 
Using Eqs.~\eqref{eq:self-consistent} and~\eqref{eq:vertex},
we find the following result for the shear viscosity at $\omega=0$ \cite{SM}:
\begin{gather}
\eta_{s}
= \frac{1}{2} \int d\varepsilon \bigl ( - f^\prime_\varepsilon \bigr )
\frac{\nu(\varepsilon) \varepsilon^2  \tau_{\rm tr,2}(\varepsilon) }{1+4 \omega_c^2\tau_{\rm tr,2}^{2}(\varepsilon) } ,
\label{eq:etaSh}
\end{gather}
where $\tau_{\rm tr,2}(\varepsilon) = \tau_{\rm tr,2} \nu_0/\nu(\varepsilon)$ is the renormalized second transport time and $1/\tau_{\rm tr,2}=1/\tau_0-1/\tau_2$ is the second transport rate at $B=0$. We note that for $k_Fd\gg 1$ 
the second transport time becomes $\tau_{\rm tr,2} = \tau_0 (k_Fd/2)^2  \gg \tau_0$. We mention that Eq.~\eqref{eq:etaSh} is analogous to the result for the dissipative conductivity~\cite{DMP}.

In the regime of overlapping LLs, $\omega_c\tau_0\ll 1$, the shear viscosity exhibits Shubnikov-de Haas-type oscillations:
\begin{equation}
\eta_{s} = 
\frac{1}{2}\frac{\nu_0 \mu^2 \tau_{\rm tr,2}}{1+
4\alpha^{2}}
\left ( 1 - \frac{16 \alpha^{2}\delta }{1+4 \alpha^{2}} 
\mathcal{F}_T
\cos\frac{2\pi \mu}{\omega_c}\right ),
\end{equation}
where $\alpha=\omega_c\tau_{\rm tr,2}$ and $\mathcal{F}_T = (2\pi^2T/\omega_c)/\sinh(2\pi^2T/\omega_c)$.
The non-oscillatory term in $\eta_{s}$ reproduces the classical result for the shear viscosity of an electron gas~\cite{Steinberg}. 

\begin{figure}[t]
\centerline{\includegraphics[width=0.36\textwidth]{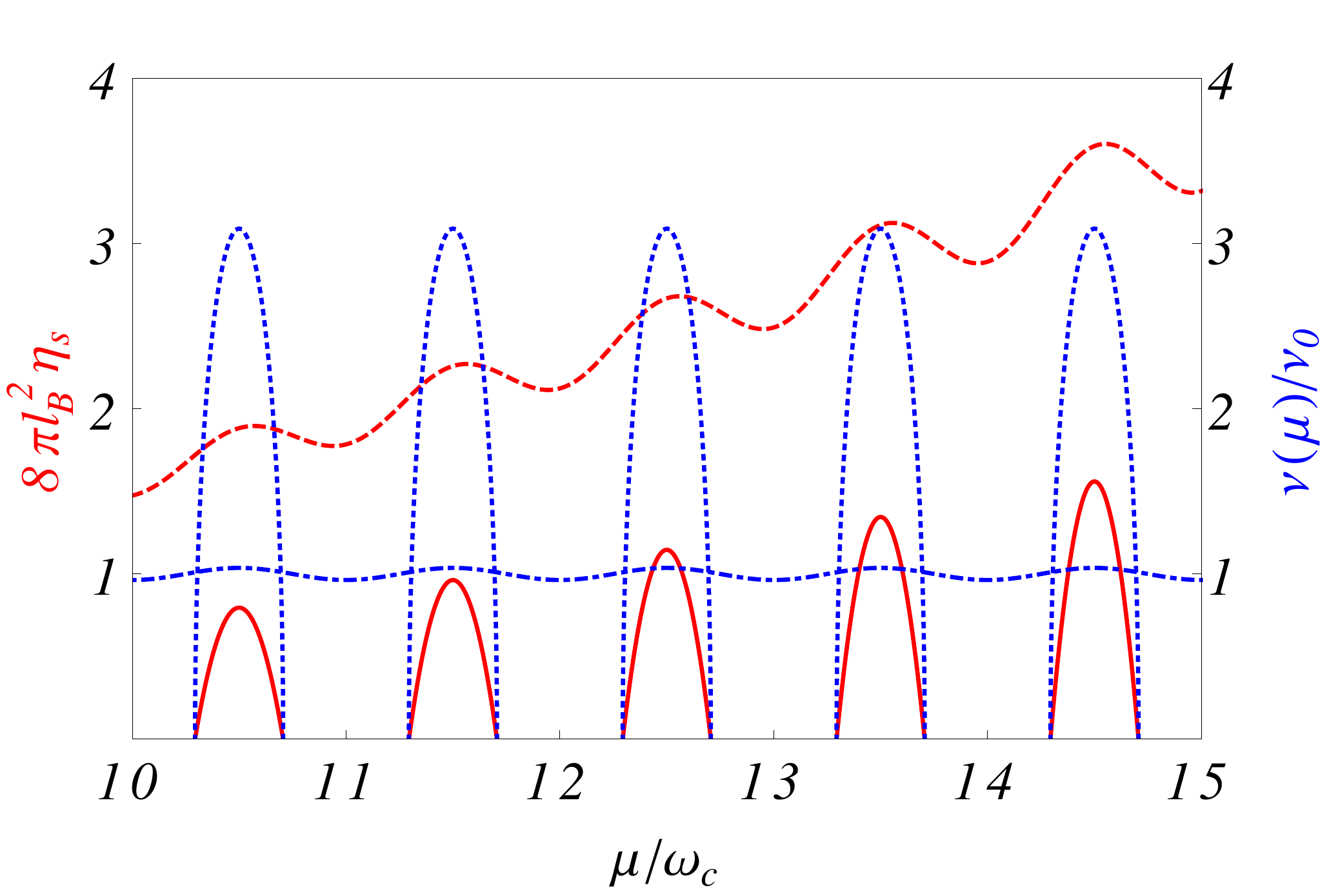}}
\caption{The density of states (blue dotted and blue dash-dotted curves) and shear viscosity (red solid and red dashed curves) as functions of $\mu/\omega_c$ for smooth disorder, $\tau_{\rm tr,2}/\tau_0=40$. Blue dotted and red solid curves correspond to well-separated LLs with $\omega_c\tau_0 = 15$. Blue dash-dotted and red dashed curves correspond to overlapping LLs with $\omega_c\tau_0=0.8$. 
}
\label{Figure-1}
\end{figure}

In the regime of well separated LLs, $\omega_c \tau_0 \gg 1$, one finds from Eq.~\eqref{eq:etaSh} that the shear viscosity is non-zero when the chemical potential is inside the $N$-th broadened Landau level ($|\mu-\epsilon_N|\leqslant \Gamma$):
\begin{equation}
\eta_{s} = (N^2 \tau_0)/(8\pi^2 l_B^2\tau_{\rm tr,2}) 
\bigl [ 1-(\mu-\epsilon_N)^2/\Gamma^2\bigr ] .
\label{eq:etas-WSLL}
\end{equation}
For chemical potential at the center of the LL, the shear viscosity is $2\omega_c\tau_0/\pi$ times larger  when one naively expects on the basis of purely classical expression.
The dependence of the shear viscosity on the chemical potential in comparison with $\nu(\varepsilon)$ is shown in Fig.~ \ref{Figure-1}.

\textsf{Hall viscosity.}\ --- The Hall viscosity can be extracted from the viscosity tensor as $\eta_H = (\eta_{xy,xx}-\eta_{xy,yy})/2$. Similarly to the Hall conductance, the evaluation of $\eta_H$ from the Kubo formula \eqref{eq:Kubo} is complicated due to contributions which come from all the states below the chemical potential. Therefore, it is convenient to proceed in a way pioneered by Smr\v{c}ka and St\v{r}eda~\cite{SmSt}: As for the Hall conductivity we split the expression for the Hall viscosity at $\omega=0$ into two parts, $\eta_H = \eta_H^I+\eta_H^{II}$, where~\cite{SM} 
\begin{align}
\eta_{H}^I = &
 \re
\int \frac{d\varepsilon}{4\pi S}  \bigl (-f^\prime_\varepsilon\bigr )
\overline{\Tr  T_{xy}  G^R_\varepsilon
\bigl (T_{xx}-T_{yy}\bigr ) G^A_\varepsilon } ,
\label{eq:etaH-Ig}
\\
\eta_H^{II}  = & \im \int \frac{d\varepsilon}{2\pi S}
 \bigl (- f^\prime_\varepsilon\bigr )
  \overline{\Tr J_{xy} (T_{yy}-T_{xx}) G^{A}_\varepsilon} \notag \\
& +  \im \int \frac{d\varepsilon}{2\pi S} f_\varepsilon \overline{\Tr [J_{xy},J_{yy}-J_{xx}]\im G^R_\varepsilon } .
\label{eq:etaH-IIg}
\end{align}
Here $J_{jk}$ denotes the \color{black} strain \color{black}
generators which are related with the stress tensor as $T_{jk}= - i [H,J_{jk}]$~\cite{Bradlyn2012}. One can evaluate $\eta_H^I$ in a similar way to $\eta_{s}$~\cite{SM}:
\begin{equation}
\eta_H^{I} = \int d\varepsilon \bigl ( - f^\prime_\varepsilon \bigr )
\frac{\nu(\varepsilon) \varepsilon^2 \omega_c\tau^2_{\rm tr,2}(\varepsilon) }{1+4 \omega_c^2\tau_{\rm tr,2}^{2}(\varepsilon) }.
\label{eq:etaH-I}
\end{equation}

The evaluation of $\eta_H^{II}$ is more involved. Although one can write down the viscoelastic analog of the 
Smr\v{c}ka and St\v{r}eda formula for the Hall viscosity \cite{VE}, it does not provided a suitable way for the calculation of $\eta_H^{II}$ in the presence of disorder. In order to compute $\eta_H^{II}$ one needs to know the expressions for the 
\color{black} strain \color{black} generators. In the absence of disorder they can be easily written down explicitly \cite{Bradlyn2012}, e.g., $J^{(0)}_{xy} = (T_{xx}-T_{yy})/(4\omega_c)$ and $J_{yy}^{(0)}-J_{xx}^{(0)} = T_{xy}/\omega_c$. In the presence of a random potential the 
\color{black} strain \color{black} generators can be constructed as a series in spatial derivatives of  a random potential $V$. This allows us to evaluate $\eta_H^{II}$ within the SCBA~\cite{SM}:
\begin{equation}
\eta_H^{II} = \mathcal{E}/(2\omega_c) - \int d\varepsilon \bigl ( - f^\prime_\varepsilon \bigr )
\nu(\varepsilon) \varepsilon^2 /(4 \omega_c) ,
\label{eq:etaH-II}
\end{equation}
where $\mathcal{E} = \int d\varepsilon \nu(\varepsilon) \varepsilon f_\varepsilon $ stands for the energy density. Combining Eqs. \eqref{eq:etaH-I} and \eqref{eq:etaH-II}, we obtain 
\begin{gather}
\eta_H = \frac{\mathcal{E}}{2\omega_c} -\frac{1}{4\omega_c} \int d\varepsilon \bigl (- f^\prime_\varepsilon\bigr )
\frac{\nu(\varepsilon) \varepsilon^2  }{1+
4\omega^2_c\tau_{\rm tr,2}^{2}(\varepsilon)} .
\label{eq:etaH}
\end{gather}
In the absence of disorder and for the chemical potential above the $N$-th Landau level the energy density at $T=0$ can be computed as $\mathcal{E}= \sum_{n=0}^N \epsilon_n/(2\pi l_B^2)$, which yields the known result
$\eta_H = \sum_{n=0}^N (n+1/2)/(4\pi l_B^2)$~\cite{Avron}. Also, we mention that in the Boltzmann limit, $T\gg E_F$, the energy density is given by $\mathcal{E} = n_e T$, where $n_e$ denotes the particle density, such that the Hall viscosity in the absence of disorder and at $T\gg E_F$ becomes $\eta_H = {n_e T}/{(2\omega_c)}$, in agreement with Eq.~(59.38) of Ref.~\cite{LL10} in which the Hall viscosity is denoted by $\eta_3$. We note that the structure of Eq.~\eqref{eq:etaH} resembles the structure of the SCBA result for the Hall conductivity $\sigma_H$~\cite{DMPZ}.

The appearance of the non-zero $\eta_H$ can be explained on a pure classical level~\cite{Kaufman}. Hall viscosity describes the response of $T_{xx}-T_{yy}$ to a shear velocity profile, $U_x = u y$. In the presence of a magnetic field this velocity can be considered as the result of a non-uniform electric field, $E_y = - U_x B$. This electric field results not only in a drift of the cyclotron orbit but in its deformation into an ellipse. To linear order in $u$ the eccentricity of the elipse is equal to $u/(2\omega_c)$. This asymmetry between motion in the $x$ and $y$ direction yields the non-zero ratio $(T_{xx}-T_{yy})/u$ in the limit $u\to 0$. Hence, non-zero $\eta_H$ arises, which is given by the first term in Eq. \eqref{eq:etaH}. An electron moving along an ellipse conserves its energy to the first order in $u$, in agreement with non-dissipative nature of $\eta_H$. In the presence of impurity scattering an electron experiences a friction force corresponding to an electric field $E_x= -U_x/(e \tau_{{\rm tr},2})$. This electric field leads to a velocity component $U_y = U_x/(\omega_c \tau_{{\rm tr},2})$. The non-uniformity of this velocity produces additional correction to the difference, $T_{xx}-T_{yy} \sim - u \eta_s/(\omega_c \tau_{{\rm tr},2})$. Thus there is an additional correction to the Hall viscosity,
$\Delta \eta_H = - \eta_s/(\omega_c \tau_{{\rm tr},2})$, which corresponds to the second term in Eq.~\eqref{eq:etaH} in the classical regime.    

\begin{figure}[t]
\centerline{\includegraphics[width=0.36\textwidth]{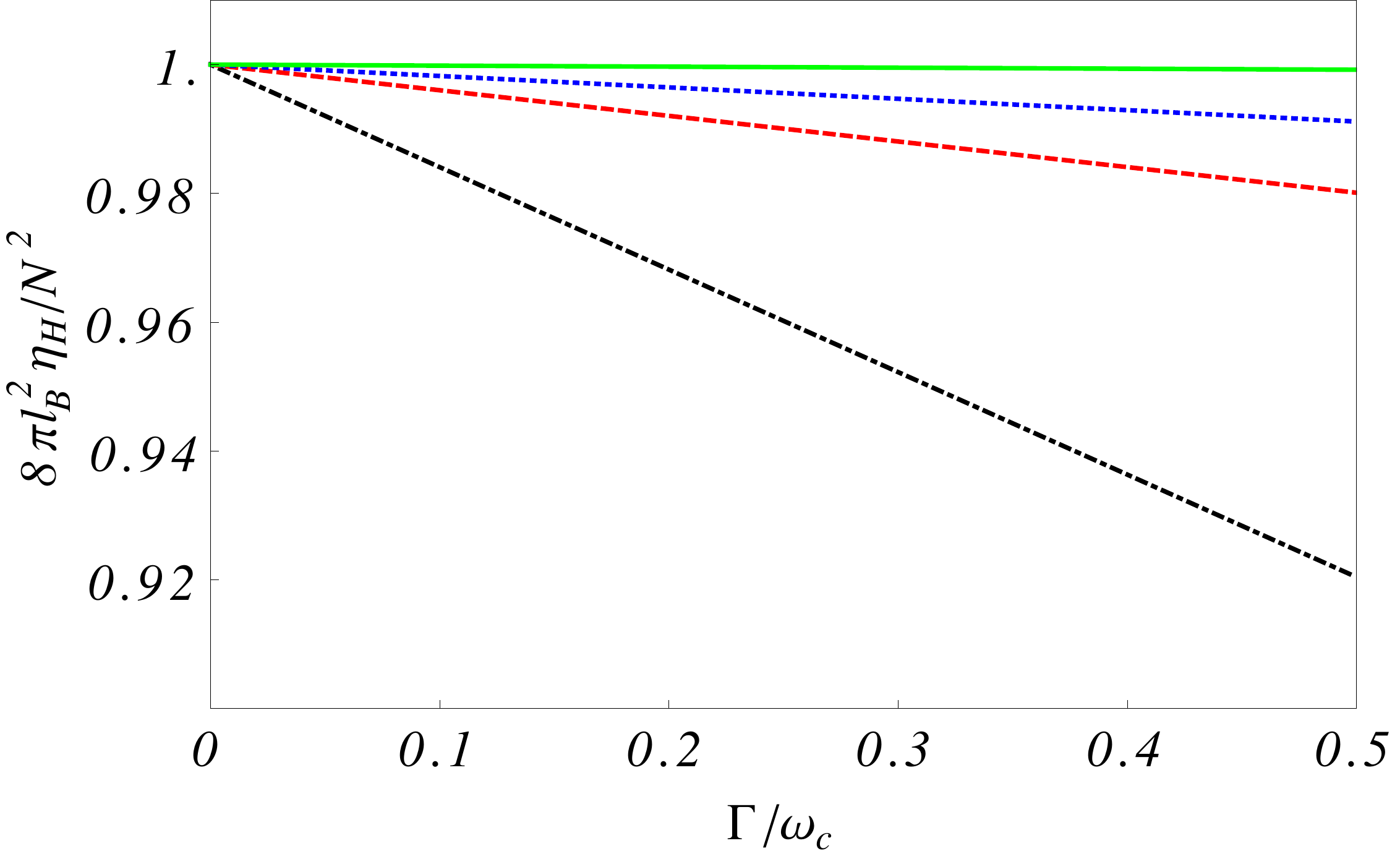}}
\caption{The normalized Hall viscosity, $8\pi l_B^2\eta_H/N^2$, as the function of $\Gamma/\omega_c$ for different ranges of disorder in the case of well-separated LLs. The parameter $\tau_{\rm tr,2}/\tau_0$ is equal to $1$, $2$, $3$, and $10$ from the bottom to the top. 
}
\label{Figure-2}
\end{figure}

In the case of overlapping LLs, $\omega_c\tau_0\ll 1$, from Eq.\eqref{eq:etaH} we obtain the Shubnikov-de Haas oscillations of the Hall viscosity: 
\begin{equation}
\eta_{\rm H} = 
\frac{\nu_0 \mu^2 \omega_c\tau_{\rm tr,2}^2}{1+
4\alpha^{2}}
\left ( 1 + \frac{\delta}{2\alpha^{2}}\frac{1+12 \alpha^{2}}{1+4 \alpha^{2}} \mathcal{F}_T
\cos\frac{2\pi \mu}{\omega_c}\right ).
\end{equation}
The non-oscillatory term in $\eta_{\rm H}$ coincides with the classical result for the Hall viscosity of electron gas \cite{Steinberg}. 

In the case of well-separated LLs, $\omega_c \tau_0 \gg 1$, one finds from Eq.~\eqref{eq:etaH} that the Hall viscosity is reduced from the quantized value if the chemical potential lies within the broadened LL, $|\mu-\epsilon_N| \leqslant \Gamma$:
\begin{equation}
\eta_H = \frac{N^2}{8\pi l_B^2} 
\left [ 
1 - 
\frac{\tau_0^2\Gamma}{2\pi\omega_c\tau_{\rm tr,2}^2}
\left (1-\frac{(\mu-\epsilon_N)^2}{\Gamma^2}\right )^{3/2}
\right ] .
\label{eq:etaH-WSLL}
\end{equation} 
We note that for the long-range-correlated random potential the Hall viscosity dominates the shear viscosity, $\eta_H\gg \eta_s$ (cf.\ Eqs.~\eqref{eq:etas-WSLL} and~\eqref{eq:etaH-WSLL}). 

The deviation of the Hall viscosity from the clean value is 
controlled by the small parameter $(\tau_0/\tau_{\rm tr,2})^2/\sqrt{\omega_c\tau_0} \ll 1$. In the case of short range random potential correlations, $\tau_0=\tau_{\rm tr,2}$, the deviation of $\eta_H$ from its clean value is very small. For long-range-correlated random potential,  $\tau_0 \ll \tau_{\rm tr,2}$, the difference $\eta_H - N^2/(8\pi l_B^2)$ is additionally suppressed (see Fig.~\ref{Figure-2}).

\begin{figure}[t]
	\centerline{\includegraphics[width=0.41\textwidth]{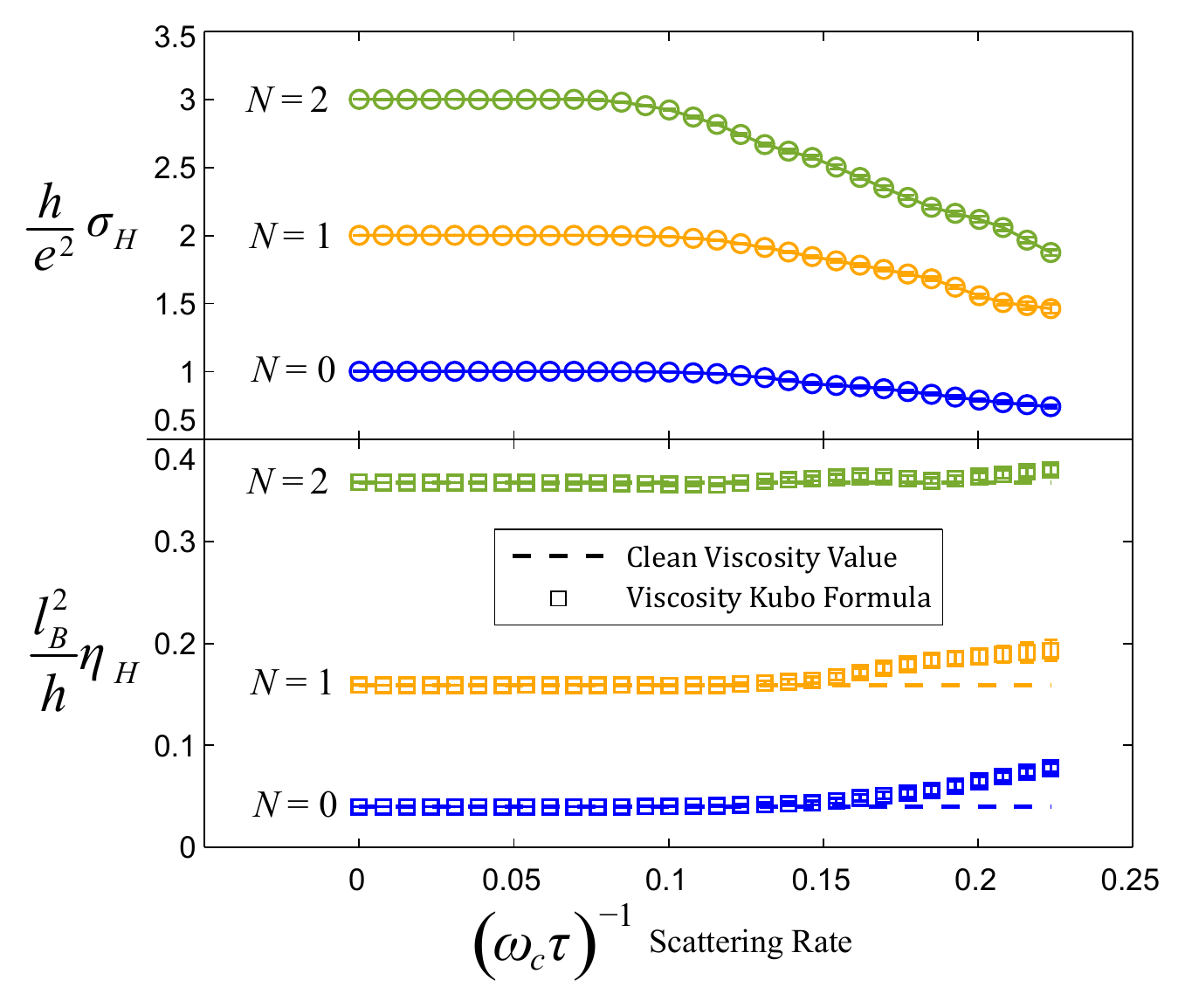}}
	\caption{The Hall conductivity (circles) and viscosity (squares) as function of $\omega_c\tau$ for $N=0, 1$, and $2$  filled Landau levels (blue, yellow, and green, respectively). The clean viscosity values are also indicated by dashed lines. $\eta_H$ is seen to be robust to disorder to the same extent the $\sigma_H$ is, within the numerical errors (error bars are smaller than the symbol sizes.)
	}
	\label{fig:numerical}
\end{figure}

\textsf{Numerical results.}\ --- 
We would now like to explore the quantum Hall regime, where the number of filled LLs is of order unity. Here the SCBA cannot be used anymore, and we resort to a numerical calculation. For this we discretize the system and employ the Hofstadter model with uncorrelated random potential, uniformly distributed between $[-w/2,w/2]$ at each lattice site. We calculate the Hall viscosity \color{black}
at zero temperature
\color{black}
using retarded correlation function of discretized stress operators~\cite{TuegelHueghes}, and take both the continuum and thermodynamic limits to extrapolate to the behavior of our model~\eqref{eq:model}~\cite{SM}. In the presence of disorder we can take these limits while keeping constant $\omega_c\tau$. The results for the Hall viscosity are plotted in Fig.~\ref{fig:numerical}, together with the behavior of the Hall conductivity ($\sigma_H$) at zero wavevector. One sees that, somewhat surprisingly, the Hall viscosity maintains its quantization to the same extent as the Hall conductivity, \color{black}
that is, until the quantum Hall to insulator transition is approached.
\color{black}

\textsf{Conclusions.}\ ---  To summarize, we studied the dissipative and Hall viscosity of 2D electron system in the presence of a random potential. Within the self-consistent Born approximation 
we derived an expressions for both the dissipative and Hall viscosities, which takes into account the modification of the single-particle density of states and the elastic transport time due to the Landau quantization. Our results smoothly interpolate between the case of weak magnetic field and strong disorder, on the one hand, and
the case of strong magnetic field and vanishing disorder, on the other hand. In the former regime, we derived the expressions for the quantum (Shubnikov-de Haas type) oscillations of the dissipative and Hall viscosities. In the case of strong magnetic field, we found that the disorder broadening of the Landau level does not lead to a significant change of the Hall viscosity in comparison with the clean result. Our numerical results for a few filled LLs support this striking conclusion. 

There are various ways to extend our work. In Galilean invariant systems it was proven~\cite{Hoyos,Bradlyn2012} that the viscosity tensor can be extracted from the nonlocal conductivity, that is, the conductivity tensor at finite wave-vector $\bm{q}$. In the absence of Galilean invariance there is no reason to expect that $\eta_H$ is related to $\sigma_H(\bm{q})$~\cite{Radzihovsky,HollerRead}. Also, the relation between $\eta_H$ and $\sigma_H(\bm{q})$ can be affected by the presence of a lattice~\cite{TuegelHueghes,Harper2018} or disorder. However, if one treats disorder on the level of Drude model with classical magnetic field, the relation of Ref.~\cite{Bradlyn2012} between $\eta_H$ and $\sigma_H(\bm{q})$ still holds \cite{HKO}. This fact is not surprising since the Drude model does not properly take into account the LLs, which result in the energy dependence of the density of states and elastic scattering transport time. However, such a simplification can be dangerous since 
$\eta_H$ and $\sigma_H$ have contributions coming from the states well below the Fermi energy. It would therefore be worthwhile to extend the presented analytical and numerical approaches to the conductivity at finite wave vector \cite{elsewhere}.
We also note that our techniques can be applied to calculation of the dissipative and Hall viscosity in graphene, where only the result in the absence of disorder is known~\cite{SPV}.

\textsf{Acknowledgements.}\ ---
We thank A. Abanov, O. Andreev, I. Gornyi, A. Mirlin, D. Polyakov, and P. Wiegmann for useful discussions. Hospitality by Tel Aviv University, the Weizmann Institute of Science, the Landau Institute for Theoretical Physics, and the Karlsruhe Institute of Technology is gratefully acknowledged. The work was partially supported by the Russian Foundation for Basic Research under Grant No.~17-02-00541, the program ``Contemporary problems of low-temperature physics'' of Russian Academy of Science, the Israel Ministry of Science and Technology (Contract No.~3-12419), the Israel Science Foundation (Grant No.~227/15), the German Israeli Foundation (Grant No.~I-1259-303.10), the US-Israel Binational Science Foundation (Grant No.~2016224), and a travel grant by the BASIS Foundation.




\section*{Supplementary material}

# Dissipative and Hall viscosity of 2D disordered electron gas

Igor S. Burmistrov,[1,2] Moshe Goldstein,[3] Mordecai Kot,[3] Vladislav D. Kurilovich,[4] and Pavel D. Kurilovich[4]

[1]*L. D. Landau Institute for Theoretical Physics, acad. Semenova av. 1-a, 142432 Chernogolovka, Russia*
[2]*Laboratory for Condensed Matter Physics, National Research University Higher School of Economics, 101000 Moscow, Russia*
[3]*Raymond and Beverly Sackler School of Physics and Astronomy, Tel Aviv University, Tel Aviv 6997801, Israel*
[4]*Departments of Physics and Applied Physics, Yale University, New Haven, CT 06520, USA*

In this notes, we present the details of (i) the definition of the stress and viscosity tensors in the presence of a random potential, (ii) the vertex renormalization in the self-consistent Born approximation, (iii) the calculation of the shear viscosity, (iv) the calculation of the Hall viscosity, (iv) the derivation of the result for the Hall viscosity in the case of well-separated Landau levels, (v) our numerical procedure for the limit of few filled Landau levels.

## S.I. THE STRESS AND VISCOSITY TENSORS IN THE PRESENCE OF A RANDOM POTENTIAL

In order to derive the form of the stress tensor in the presence of a random potential we shall follow approach of Ref. [S1]. The continuity equation for the kinetic momentum density, $g_\nu(\boldsymbol{x}) = \sum_j \{\pi_\nu^{(j)}, \delta(\boldsymbol{x} - \boldsymbol{x}^{(j)})\}/2$, has the following form [S1, S2]:

$$\frac{\partial g_\nu(\boldsymbol{x})}{\partial t} + \partial_\mu \tau_{\mu\nu}(\boldsymbol{x}) = \frac{1}{2}\sum_j \{f_\nu^{(j)}, \delta(\boldsymbol{x} - \boldsymbol{x}^{(j)})\}, \qquad f_\nu^{(j)} = eB\epsilon_{\nu\mu} v_\mu^{(j)} - \frac{\partial V(\boldsymbol{x}^{(j)})}{\partial x_\nu^{(i)}}. \tag{S1}$$

Here $\tau_{\mu\nu}(\boldsymbol{x})$ stands for the stress tensor density. $\boldsymbol{\pi}^{(j)} = m_e \boldsymbol{v}^{(j)}$ and $\boldsymbol{x}^{(j)}$ denote, respectively, the kinetic momentum and coordinate operators of the $j$-th particle. Introducing $T_{\mu\nu} = \int d^2\boldsymbol{x}\, \tau_{\mu\nu}$ and taking the Fourier transform, we obtain for $q \to 0$:

$$\frac{\partial}{\partial t}\sum_j \left[\pi_\nu^{(i)} - \frac{iq_\mu}{2}\{x_\mu^{(j)}, \pi_\nu^{(j)}\}\right] + iq_\mu T_{\mu\nu} = \sum_j \left[f_\nu^{(j)} - \frac{iq_\mu}{2}\{x_\mu^{(j)}, f_\nu^{(j)}\}\right]. \tag{S2}$$

Here the terms independent of $q$ describes the change of the total momentum due to the Lorentz force and the force produced by a random potential. The terms proportional to $q$ determine the form of the stress tensor:

$$T_{\mu\nu} = \frac{1}{2}\sum_j \left(\frac{\partial}{\partial t}\{x_\mu^{(j)}, \pi_\nu^{(j)}\} - \{x_\mu^{(j)}, f_\nu^{(j)}\}\right) = \frac{1}{2}\sum_j \left(i[H, \{x_\mu^{(j)}, \pi_\nu^{(j)}\}] - \{x_\mu^{(j)}, f_\nu^{(j)}\}\right) \equiv \frac{m_e}{2}\sum_j \{v_\mu^{(j)}, v_\nu^{(j)}\}. \tag{S3}$$

We see that the form of the stress tensor is insensitive to the presence of a random potential, since the disorder enters into the non-momentum-conserving force density $f_\nu$, and not the stress tensor.

Let us now consider a deformation of the Hamiltonian $H \to H_\Lambda$ under the application of a strain $\Lambda$ that transforms the coordinates as $x_\mu \to \Lambda_{\nu\mu} x_\nu$. We note that the momentum should be transformed as $\pi_\mu \to (\Lambda^{-1})_{\mu\nu}\pi_\nu$ in order to preserve the commutation relation $[x_\mu, \pi_\nu] = i\delta_{\mu\nu}$. The expansion of the deformed Hamiltonian $H_\Lambda$ to first order in the matrix $\lambda = \ln\Lambda$ determines the stress tensor: $H_\Lambda = H - \lambda_{\mu\nu} T_{\mu\nu} + O(\lambda^2)$. This definition of the stress tensor is consistent with Eq. (S3) provided the deformed Hamiltonian has the following form:

$$H_\Lambda = \frac{1}{2m_e}(\Lambda^{-1})_{\alpha\mu}(\Lambda^{-1})_{\alpha\nu}\pi_\mu\pi_\nu + V(\boldsymbol{x}). \tag{S4}$$

Eq. (S4) indicates that strains do not affect a random potential. Physically this means that one strains the electron liquid rather than the host material. This is consistent with analysis of Ref. [S1] (especially Appendix C) for the case of a confining potential, as well as with the approach based on the effective action for electrons in a time dependent metric [S3], which does not couple to a scalar potential.

Let us now introduce the strain generator $J_{\mu\nu}$ as $T_{\mu\nu} = -i[H, J_{\mu\nu}]$. We emphasize that the form of the strain generator is affected by a random potential, since $H$ is and $T_{\mu\nu}$ is not. A general formalism for its construction in the presence of a potential is given in Appendix B of Ref. [S1], though for our purposes we employ an expansion in the spatial derivatives of the random potential, Eq. (S40) below. With the strain generators Eq. (S4) can be written as $H_\Lambda = SHS^{-1}$ where $S = \exp(-i\lambda_{\mu\nu} J_{\mu\nu})$. For a time dependent strain $\Lambda(t)$ we transform the Hamiltonian $H_\Lambda$ into

$H_\Lambda \to H - iS\partial S^{-1}/\partial t = H - J_{\mu\nu}\partial \lambda_{\mu\nu}/\partial t$. Thus, $J_{\mu\nu}$ governs the time-dependent deformation perturbation, giving rise to the stress-strain form of the Kubo formula for the viscosity [S1]. Using the relation $T_{\mu\nu} = -i[H, J_{\mu\nu}] = -\partial J_{\mu\nu}/\partial t$, the Kubo formula can be transformed into the stress-stress form which we use in the main text [Eq. (2)].

## S.II. THE SELF-CONSISTENT BORN APPROXIMATION: THE SELF ENERGY AND VERTEX RENORMALIZATION

In this section we present details of derivation for the self energy in the average Green's function and the vertex renormalization within the self-consistent Born approximation (SCBA) [S4]. As in the main text, in order to justify the SCBA, we assume that the following conditions hold [S5, S6, S7]:

$$1/k_F, d \ll l_B, \qquad d \ll v_F \tau_0. \tag{S5}$$

We work in the Landau gauge, $A_x = A_z = 0$ and $A_y = -Bx$. The corresponding basis of the eigenfunctions of the 2D Landau problem is denoted by $|nk\rangle$, where $n$ enumerates the Landau levels (LLs), whereas $k$ denotes the pseudo-momentum. In what follows, we shall use the following identities:

$$\langle nk|e^{i\bm{q r}}|n'k'\rangle = \frac{2\pi}{L_y}\delta(k-k'-q_y)e^{iq_x(k+k')l_B^2/2}\sqrt{\frac{n'!}{n!}}(-i)^{n-n'}e^{-q^2l_B^2/4}\left(\frac{q_- l_B}{\sqrt{2}}\right)^{n-n'} L_{n'}^{n-n'}\left(\frac{q^2 l_B^2}{2}\right)$$

$$= \frac{2\pi}{L_y}\delta(k-k'-q_y)e^{iq_x(k+k')l_B^2/2}\sqrt{\frac{n!}{n'!}}(-i)^{n'-n}e^{-q^2l_B^2/4}\left(\frac{q_+ l_B}{\sqrt{2}}\right)^{n'-n} L_n^{n'-n}\left(\frac{q^2 l_B^2}{2}\right), \tag{S6}$$

where $q_\pm = q_x \pm i q_y$ and $L_n^m(z)$ denotes a generalized Laguerre polynomial.

The equation for the self-consistent Green's function can be written as ($\epsilon_n = \omega_c(n+1/2)$)

$$\mathcal{G}_n^R(\varepsilon) = [\varepsilon - \epsilon_n - \Sigma_n^R(\varepsilon)]^{-1},$$

$$\Sigma_n^R(\varepsilon) = \int \frac{d^2\bm{Q}}{(2\pi)^2} W(Q) \sum_{n'} \frac{n'!}{n!} \mathcal{G}_{n'}^R(\varepsilon) e^{-z_Q} z_Q^{n-n'} L_{n'}^{n-n'}(z_Q) L_{n'}^{n-n'}(z_Q). \tag{S7}$$

Here and afterwards, $z_Q = Q^2 l_B^2/2$. The Fourier transform of the pair correlation function of the random potential is denoted by $W(Q)$. To simplify the expression for the self-energy we can use the following asymptotic form of the Laguerre polynomials ($n \gg a$):

$$L_n^a(x) = \sqrt{\frac{(n+a)!}{n!}} e^{x/2} x^{-a/2} J_a\left(\sqrt{2x(2n+a+1)}\right), \tag{S8}$$

where $J_a(x)$ stands for the Bessel function. We thus find the following expression for the self-energy

$$\Sigma_n^R(\varepsilon) = \sum_{n'} \mathcal{G}_{n'}^R(\varepsilon) \int \frac{d^2\bm{Q}}{(2\pi)^2} W(Q) J_{n-n'}^2\left(Q l_B \sqrt{n+n'+1}\right) \approx \int\limits_0^\infty \frac{dQ}{2\pi^2 R_c} W(Q) \sum_{n'} \mathcal{G}_{n'}(\varepsilon), \tag{S9}$$

where we used the asymptotic form of the Bessel function, relying on the condition $QR_c \gg 1$, where $R_c = l_B\sqrt{n+n'+1}$ is the cyclotron radius. Provided that only the LLs which are close to the energy $\varepsilon$ contribute significantly, we can approximate the cyclotron radius as $R_c = v_\varepsilon/\omega_c$, where $v_\varepsilon = \sqrt{2\varepsilon/m}$. Then the self-energy becomes independent of $n$. Using the expression for the total scattering rate at zero magnetic field

$$\frac{1}{\tau_0} = 2\pi\nu_0 \int\limits_0^{2\pi} d\phi\, W\left(2k_\varepsilon|\sin(\phi/2)|\right) \approx \frac{1}{\pi v_\varepsilon} \int\limits_0^\infty dQ\, W(Q) = \frac{2\pi}{\omega_c} \int\limits_0^\infty \frac{dQ}{2\pi^2 R_c} W(Q), \tag{S10}$$

where $\nu_0 = m_e/(2\pi)$ and $k_\varepsilon = m_e v_\varepsilon$, we obtain the following self-consistent equation for the self energy:

$$\Sigma_\varepsilon^R = \frac{\omega_c}{2\pi\tau_0} \sum_{n'} \mathcal{G}_{n'}^R(\varepsilon). \tag{S11}$$

This equation demonstrates that the self-energy is related to the density of states as $\mathrm{Im}\,\Sigma^R(\varepsilon) = -\nu(\varepsilon)/(2\nu_0\tau_0)$.

Within the SCBA one can use the quasiclassical form of the matrix elements for the components of the stress tensor:

$$T_{xy} = -\frac{i\varepsilon}{2}(V_2 - V_{-2}), \quad T_{xx} - T_{yy} = -\varepsilon(V_2 + V_{-2}), \qquad \langle nk|V_m|n'k'\rangle = \delta_{k,k'}\delta_{n',n+m}. \tag{S12}$$

In order to evaluate the viscosity tensor one needs to know the renormalization of the operator $V_m$ by the ladder of disorder lines. The corresponding renormalized vertex $\Gamma_m$ is diagonal in the pseudo-momentum $k$ and satisfies the following equation:

$$\Gamma_m^{RA} = V_m + \sum_{n_1} \left[\frac{(n+m)!n!}{(n_1+m)!n_1!}\right]^{1/2} \mathcal{G}_{n_1+m}^R(\varepsilon+\omega)\mathcal{G}_{n_1}^A(\varepsilon)\Gamma_m^{RA} \int \frac{d^2\boldsymbol{Q}}{(2\pi)^2} W(Q) e^{-z_Q} z_Q^{n_1-n} L_n^{n_1-n}(z_Q) L_{n+m}^{n_1-n}(z_Q). \tag{S13}$$

Similar equations hold for $\Gamma_m^{AR}$, $\Gamma_m^{RR}$ and $\Gamma_m^{AA}$. Using the asymptotic expression (S8) we find

$$\Gamma_m^{RA} = V_m + \sum_{n_1} \mathcal{G}_{n_1+m}^s(\varepsilon+\omega)\mathcal{G}_{n_1}^{s'}(\varepsilon) \int \frac{d^2\boldsymbol{Q}}{(2\pi)^2} W(Q) J_{n_1-n}(Ql_B\sqrt{n+n_1+1}) J_{n_1-n}(Ql_B\sqrt{n+n_1+1+2m})\Gamma_m^{RA}. \tag{S14}$$

It is convenient to introduce the quantity

$$\frac{1}{\tau_m} = \frac{2\pi}{\omega_c} \int \frac{d^2\boldsymbol{Q}}{(2\pi)^2} W(Q) J_{n_1-n}(Ql_B\sqrt{n+n_1+1}) J_{n_1-n}(Ql_B\sqrt{n+n_1+1+2m}) \tag{S15}$$

Using the asymptotic expressions for the Bessel functions at $QR_c \gg 1$, we find

$$\frac{1}{\tau_m} = \frac{1}{\pi\omega_c R_c} \int_0^\infty dQ\, W(Q) \cos\left[Q\sqrt{R_c^2 + 2ml_B^2} - QR_c\right] \approx \frac{1}{\tau_0} - \frac{m^2}{2\pi v_\varepsilon} \int_0^\infty dQ (Q/k_\varepsilon)^2 W(Q). \tag{S16}$$

We note that the transport elastic rate is given as $1/\tau_{\text{tr},1} = 1/\tau_0 - 1/\tau_1$. Finally, we obtain

$$\Gamma_m^{RA} = \frac{V_m}{1 - \Pi_m^{RA}(\omega)/\tau_m}, \qquad \Pi_m^{RA}(\omega) = \frac{\omega_c}{2\pi}\sum_{n_1} \mathcal{G}_{n_1+m}^R(\varepsilon+\omega)\mathcal{G}_{n_1}^A(\varepsilon) = \frac{\tau_0(\Sigma_{\varepsilon+\omega}^R - \Sigma_\varepsilon^A)}{m\omega_c - \omega + \Sigma_{\varepsilon+\omega}^R - \Sigma_\varepsilon^A}. \tag{S17}$$

## S.III. EVALUATION OF THE SHEAR VISCOSITY

Using the Kubo formula from the main text we can write the real part of the shear viscosity as

$$\begin{aligned}
\operatorname{Re}\eta_s(\omega) &= \operatorname{Im}\int \frac{d\Omega}{\Omega-\omega-i0}\int \frac{d\varepsilon}{\pi^2 S}\frac{f_\varepsilon - f_{\varepsilon+\Omega}}{\omega}\, \overline{\operatorname{Tr} T_{xy}\operatorname{Im} G_{\varepsilon+\Omega}^R T_{xy}\operatorname{Im} G_\varepsilon^R}\\
&= \operatorname{Im}\int \frac{d\Omega}{\Omega-\omega-i0}\int d\varepsilon\, \frac{f_\varepsilon - f_{\varepsilon+\Omega}}{\omega}\, \overline{\sum_{a,b}|\langle a|T_{xy}|b\rangle|^2 \delta(\varepsilon+\Omega-E_b)\delta(\varepsilon-E_a)}\\
&= \pi\int d\varepsilon\, \frac{f_\varepsilon - f_{\varepsilon+\omega}}{\omega}\, \overline{\sum_{a,b}|\langle a|T_{xy}|b\rangle|^2\delta(\varepsilon+\omega-E_b)\delta(\varepsilon-E_a)}\\
&= \int \frac{d\varepsilon}{\pi S}\frac{f_\varepsilon - f_{\varepsilon+\omega}}{\omega}\,\overline{\operatorname{Tr} T_{xy}\operatorname{Im} G_{\varepsilon+\omega}^R T_{xy}\operatorname{Im} G_\varepsilon^R}\ . \tag{S18}
\end{aligned}$$

Here $E_a$, $E_b$ and $|a\rangle$, $|b\rangle$ denote the eigenenergies and eigenfunctions of $H$ for a given disorder realization. As one can see, $\eta_s$ is determined by the states which are close to the Fermi energy. Using the quasiclassical result (S12) for the stress tensor and the result (S17) for the vertex we find

$$\operatorname{Re}\eta_s(\omega) = \frac{\nu_0}{4\omega}\int d\varepsilon\, \varepsilon^2 \Big(f_\varepsilon - f_{\varepsilon+\omega}\Big) \operatorname{Re}\sum_{\sigma=\pm}\left(\frac{\Pi_{2\sigma}^{RA}(\omega)}{1-\Pi_{2\sigma}^{RA}(\omega)/\tau_2} - \frac{\Pi_{2\sigma}^{RR}(\omega)}{1-\Pi_{2\sigma}^{RR}(\omega)/\tau_2}\right). \tag{S19}$$

Setting the frequency $\omega$ to zero and using the expression (S17) for the polarization operators $\Pi_{2\sigma}^{RA}$ and $\Pi_{2\sigma}^{RR}$, we obtain the result for $\eta_s$ from the main text.

## S.IV. EVALUATION OF THE HALL VISCOSITY

### A. Splitting the Hall viscosity into $\eta_H^I$ and $\eta_H^{II}$

Using the Kubo formula from the main text we can write the Hall viscosity as

$$\eta_H(\omega) = \int \frac{d\Omega}{\Omega - \omega - i0} \int \frac{d\varepsilon}{2\pi^2 iS} \frac{f_\varepsilon - f_{\varepsilon+\Omega}}{\omega} \overline{\operatorname{Tr} T_{xy} \operatorname{Im} G_{\varepsilon+\Omega}^R (T_{xx} - T_{yy}) \operatorname{Im} G_\varepsilon^R} . \tag{S20}$$

Following Smrčka and Středa [S8], we split the Hall viscosity into two parts $\eta_H(\omega) = \eta_H^I(\omega) + \eta_H^{II}(\omega)$, where

$$\begin{aligned}
\eta_H^I(\omega) &= \int \frac{d\Omega}{\Omega - \omega - i0} \int \frac{d\varepsilon}{8\pi^2 iS} \frac{f_\varepsilon - f_{\varepsilon+\Omega}}{\omega} \overline{\operatorname{Tr}\left[ T_{xy} G_{\varepsilon+\Omega}^R (T_{xx} - T_{yy}) G_\varepsilon^A - T_{xy} G_{\varepsilon+\Omega}^A (T_{xx} - T_{yy}) G_\varepsilon^R \right]}, \\
\eta_H^{II}(\omega) &= \int \frac{d\Omega}{\Omega - \omega - i0} \int \frac{d\varepsilon}{4\pi^2 S} \frac{f_\varepsilon - f_{\varepsilon+\Omega}}{\omega} \overline{\operatorname{Tr}\left[ T_{xy} G_{\varepsilon+\Omega}^A (T_{xx} - T_{yy}) \operatorname{Im} G_\varepsilon^R - T_{xy} \operatorname{Im} G_{\varepsilon+\Omega}^R (T_{xx} - T_{yy}) G_\varepsilon^R \right]}.
\end{aligned} \tag{S21}$$

#### 1. Evaluation of $\eta_H^I$

It is convenient to rewrite the expression for the $\eta_H^I$ in terms of the exact eigenstates:

$$\eta_H^I(\omega) = \int \frac{d\Omega}{\Omega - \omega - i0} \int \frac{d\varepsilon}{8\pi^2 i} \frac{f_\varepsilon - f_{\varepsilon+\Omega}}{\omega} \overline{\sum_{a,b} \left[ \frac{\langle a|T_{xy}|b\rangle\langle b|(T_{xx}-T_{yy})|a\rangle}{(\varepsilon+\Omega-E_b+i0)(\varepsilon-E_a-i0)} - \frac{\langle a|T_{xy}|b\rangle\langle b|(T_{xx}-T_{yy})|a\rangle}{(\varepsilon+\Omega-E_b-i0)(\varepsilon-E_a+i0)} \right]}. \tag{S22}$$

Relations (S12) imply that one can rewrite the above expression as

$$\begin{aligned}
\operatorname{Re}\eta_H^I(\omega) &= \operatorname{Im} \int \frac{d\Omega}{\Omega - \omega - i0} \int \frac{d\varepsilon}{4\pi^2} \frac{f_\varepsilon - f_{\varepsilon+\Omega}}{\omega} \operatorname{Re} \overline{\sum_{a,b} \frac{\langle a|T_{xy}|b\rangle\langle b|(T_{xx}-T_{yy})|a\rangle}{(\varepsilon+\Omega-E_b+i0)(\varepsilon-E_a-i0)}} \\
&= \operatorname{Re} \int \frac{d\varepsilon}{4\pi S} \frac{f_\varepsilon - f_{\varepsilon+\omega}}{\omega} \overline{\operatorname{Tr} T_{xy} G_{\varepsilon+\omega}^R (T_{xx}-T_{yy}) G_\varepsilon^A}.
\end{aligned} \tag{S23}$$

Therefore, similarly to $\eta_s$, $\eta_H^I$ is determined by the states which are close to the Fermi energy. Therefore, we can compute $\eta_H^I$ in the same way as $\eta_s$. Setting $\omega$ to zero, we obtain the Kubo formula for $\eta_H^I$ given in the main text.

Finally, using the result (S17) for the renormalized vertex and the quasiclassical result (S12) for the stress tensor, we obtain

$$\operatorname{Re}\eta_H^I(\omega) = -\frac{\nu_0}{4\omega} \int d\varepsilon \left( f_\varepsilon - f_{\varepsilon+\omega} \right) \varepsilon^2 \operatorname{Im} \sum_\sigma \frac{\sigma \Pi_{2\sigma}^{RA}(\omega)}{1 - \Pi_{2\sigma}^{RA}(\omega)/\tau_2}. \tag{S24}$$

In the limit $\omega \to 0$, we reproduce the result for $\eta_H^I$ from the main text.

#### 2. Evaluation of $\eta_H^{II}$

For the evaluation of $\eta_H^{II}$ we can rewrite Eq. (S21) as

$$\begin{aligned}
\eta_H^{II}(\omega) &= \int \frac{d\Omega d\varepsilon}{4\pi\omega} f_\varepsilon \overline{\sum_{a,b} \frac{\langle a|T_{xy}|b\rangle\langle b|(T_{xx}-T_{yy})|a\rangle}{(\Omega-\omega-i0)(\Omega+E_a-E_b-i0)}} \left[ \delta(\varepsilon-E_b) - \delta(\varepsilon+\Omega-E_b) + \delta(\varepsilon-\Omega-E_a) - \delta(\varepsilon-E_a) \right] \\
&= -\int \frac{d\varepsilon}{4\pi\omega S} f_\varepsilon \overline{\operatorname{Tr}\left[ T_{xy} G_{\varepsilon+\omega}^R (T_{xx}-T_{yy}) G_\varepsilon^R - T_{xy} G_\varepsilon^A (T_{xx}-T_{yy}) G_{\varepsilon-\omega}^A \right]}.
\end{aligned} \tag{S25}$$

Taking the limit $\omega \to 0$, we obtain

$$\eta_H^{II} = \int \frac{d\varepsilon}{4\pi} f_\varepsilon \overline{\sum_{a,b} \langle a|T_{xy}|b\rangle \langle b|(T_{xx}-T_{yy})|a\rangle \left[\frac{1}{(\varepsilon - E_b + i0)^2(\varepsilon - E_a + i0)} + \frac{1}{(\varepsilon - E_b - i0)(\varepsilon - E_a - i0)^2}\right]}$$

$$= \int \frac{d\varepsilon}{4\pi} f_\varepsilon \overline{\sum_{a,b} \frac{\langle a|T_{xy}|b\rangle \langle b|(T_{xx}-T_{yy})|a\rangle}{E_b - E_a} \left[\frac{1}{(\varepsilon - E_b + i0)^2} - \frac{1}{(\varepsilon - E_a - i0)^2} + 2\pi i \frac{\delta(\varepsilon - E_a) - \delta(\varepsilon - E_b)}{E_a - E_b}\right]}$$

$$= -\operatorname{Im} \int \frac{d\varepsilon}{2\pi S}(-f'_\varepsilon)\, \overline{\operatorname{Tr} J_{xy}(T_{xx}-T_{yy})G_\varepsilon^A} - \operatorname{Im} \int \frac{d\varepsilon}{2\pi S} f_\varepsilon \overline{\operatorname{Tr}[J_{xy}, J_{xx}-J_{yy}]\operatorname{Im} G_\varepsilon^R}. \qquad (S26)$$

Now we can use the quasiclassical approximations for the strain generators:

$$J_{xy} \approx \frac{T_{xx}-T_{yy}}{4\omega_c}, \quad J_{xx} - J_{yy} \approx -\frac{T_{xy}}{\omega_c}. \qquad (S27)$$

Then, we obtain

$$\operatorname{Im} \int \frac{d\varepsilon}{2\pi S}(-f'_\varepsilon)\, \overline{\operatorname{Tr} J_{xy}(T_{xx}-T_{yy})G_\varepsilon^A} = \operatorname{Im} \int \frac{d\varepsilon}{8\pi\omega_c S}(-f'_\varepsilon)\, \overline{\operatorname{Tr}(T_{xx}-T_{yy})^2 G_\varepsilon^A}$$

$$= \operatorname{Im} \int \frac{d\varepsilon}{8\pi\omega_c}(-f'_\varepsilon) 2\varepsilon^2 \frac{1}{2\pi l_B^2} \sum_n \mathcal{G}_n^A(\varepsilon) = \int \frac{d\varepsilon}{4\omega_c}(-f'_\varepsilon)\nu(\varepsilon)\varepsilon^2. \qquad (S28)$$

Next, using the relation

$$i[T_{xx}-T_{yy}, T_{xy}] = 4\omega_c H_0, \qquad H_0 = \frac{1}{2m_e}(\boldsymbol{p} - e\boldsymbol{A})^2, \qquad (S29)$$

we find in the quasiclassical approximation:

$$\operatorname{Im} \int \frac{d\varepsilon}{2\pi S} f_\varepsilon \overline{\operatorname{Tr}[J_{xy}, J_{xx}-J_{yy}]\operatorname{Im} G_\varepsilon^R} = \operatorname{Re} \int \frac{d\varepsilon}{2\pi\omega_c S} f_\varepsilon \overline{\operatorname{Tr} H_0 \operatorname{Im} G_\varepsilon^R} = -\int \frac{d\varepsilon}{2\omega_c} f_\varepsilon \overline{\sum_a E_a \delta(\varepsilon - E_a)}$$

$$= -\frac{1}{2\omega_c}\int d\varepsilon\, \nu(\varepsilon)\varepsilon f_\varepsilon. \qquad (S30)$$

Combining together Eqs. (S28) and (S30), we obtain the result for $\eta_H^{II}$ from the main text.

### S.V. SEPARATED LANDAU LEVELS

In this section we demonstrate in details how the quasiclassical approximation works for the case of well-separated Landau levels, $\omega_c \tau_0 \gg 1$. For simplicity we consider the case of short-ranged random potential, $d \ll \lambda_F$. In this case one does not need to renormalize vertex by the disorder ladder. For short-ranged random potential the conditions (S5) are equivalent to the condition $\mu \gg \omega_c$.

We start from the evaluation of $\eta_H^I$. Using Eq. (S23) we find

$$\eta_H^I = \operatorname{Re} \int \frac{d\varepsilon}{4\pi}(-f'_\varepsilon) \sum_{n,n'} \sum_{k,k'} \langle nk|T_{xy}|n'k'\rangle \langle n'k'|T_{xx}-T_{yy}|nk\rangle \mathcal{G}_{n'}^R(\varepsilon)\mathcal{G}_n^A(\varepsilon)$$

$$= -\frac{m_e \omega_c^4}{4\pi^2} \int d\varepsilon(-f'_\varepsilon) \sum_n \frac{n(n-1)\operatorname{Im}\Sigma_\varepsilon^R}{|\varepsilon - \epsilon_n - \Sigma_\varepsilon^R|^2 |\varepsilon - \epsilon_{n-2} - \Sigma_\varepsilon^R|^2}. \qquad (S31)$$

Here we used the matrix elements of components of the stress tensor:

$$\langle nk|T_{xy}|n'k'\rangle = im_e \langle nk|v_+^2 - v_-^2|n'k'\rangle = -\frac{i}{2m_e l_B^2}\delta_{k,k'}\left(\sqrt{n'(n'-1)}\delta_{n',n+2} - \sqrt{n(n-1)}\delta_{n,n'+2}\right),$$

$$\langle nk|T_{xx}-T_{yy}|n'k'\rangle = 2m_e \langle nk|v_+^2 + v_-^2|n'k'\rangle = -\frac{1}{m_e l_B^2}\delta_{k,k'}\left(\sqrt{n'(n'-1)}\delta_{n',n+2} + \sqrt{n(n-1)}\delta_{n,n'+2}\right). \qquad (S32)$$

In order to proceed further, we need to solve the self-consistent equation (S11) for the self-energy. For well-separated Landau levels, $\omega_c \gg \Gamma$, we can solve it by expanding the right hand side in powers of $1/\omega_c$:

$$\Sigma_\varepsilon^R = \frac{\omega_c}{2\pi\tau_0} \sum_n \frac{1}{\varepsilon - \epsilon_n - \Sigma_\varepsilon^R} = \frac{\omega_c}{2\pi\tau_0} \frac{1}{\varepsilon - \epsilon_N - \Sigma_\varepsilon^R} + \frac{\omega_c}{2\pi\tau_0} \sum_{n \neq N} \frac{1}{\varepsilon - \epsilon_n - \Sigma_\varepsilon^R}$$
$$= \frac{\Gamma^2}{4} \frac{1}{\varepsilon - \epsilon_N - \Sigma_\varepsilon^R} + \frac{\Gamma^2}{4} \sum_{n \neq N} \left[ \frac{1}{\epsilon_N - \epsilon_n} - \frac{\varepsilon - \epsilon_N - \Sigma_\varepsilon^R}{(\epsilon_N - \epsilon_n)^2} + \ldots \right]. \tag{S33}$$

Here $N$ is the index of the Landau level which is the closest one to the energy $\varepsilon$, i.e. $|\varepsilon - \epsilon_N| < \omega_c/2$. We note that the conditions of applicability of SCBA implies that $N \gg 1$. Solving Eq. (S33) we obtain

$$\Sigma_\varepsilon^{R,A} \approx \frac{1}{2} \left[ (1 - \gamma)(\varepsilon + \Delta - \epsilon_N) \mp i(1 + \gamma)\sqrt{(1 - \gamma)\Gamma^2 - (\varepsilon - \Delta - \epsilon_N)^2} \right], \tag{S34}$$

where

$$\Delta = \frac{\Gamma^2}{4} \sum_{n \neq N} \frac{1}{\epsilon_N - \epsilon_n}, \qquad \gamma = \frac{\Gamma^2}{4} \sum_{n \neq N} \frac{1}{(\epsilon_N - \epsilon_n)^2}. \tag{S35}$$

We note that $\Delta \propto \Gamma^2/\omega_c$ and $\gamma \propto \Gamma^2/\omega_c^2$. The result (S34) leads to the following expression for the density of states:

$$\nu(\varepsilon) = \nu_0 \tau_0 (1 + \gamma)\tilde{\Gamma}(\varepsilon) = \frac{B}{\pi^2}(1 + \gamma)\frac{\tilde{\Gamma}(\varepsilon)}{\Gamma^2}, \qquad \tilde{\Gamma}(\varepsilon) = \sqrt{(1 - \gamma)\Gamma^2 - (\varepsilon - \Delta - \epsilon_N)^2}. \tag{S36}$$

In a similar way, expanding in powers of $1/\omega_c$ we compute the sum over LLs on the right hand side of Eq. (S31):

$$\sum_n \frac{n(n-1)}{|\varepsilon - \epsilon_n - \Sigma_\varepsilon^R|^2 |\varepsilon - \epsilon_{n-2} - \Sigma_\varepsilon^R|^2} = \frac{1}{|\varepsilon - \epsilon_N - \Sigma_\varepsilon^R|^2} \left( \frac{N(N-1)}{|\varepsilon - \epsilon_{N-2} - \Sigma_\varepsilon^R|^2} + \frac{(N+2)(N+1)}{|\varepsilon - \epsilon_{N+2} - \Sigma_\varepsilon^R|^2} \right)$$
$$+ \sum_{n \neq N, N+2} \frac{n(n-1)}{|\varepsilon - \epsilon_n - \Sigma_\varepsilon^R|^2 |E - \epsilon_{n-2} - \Sigma_\varepsilon^R|^2} \approx \frac{4(1-\gamma)}{\Gamma^2} \left\{ \frac{N(N-1)}{4\omega_c^2} \left[ 1 - \frac{\varepsilon - \epsilon_N}{2\omega_c} + \frac{(\varepsilon - \epsilon_N)^2 - \Gamma^2/4}{4\omega_c^2} \right] \right.$$
$$\left. + \frac{(N+2)(N+1)}{4\omega_c^2} \left[ 1 + \frac{\varepsilon - \epsilon_N}{2\omega_c} + \frac{(\varepsilon - \epsilon_N)^2 - \Gamma^2/4}{4\omega_c^2} \right] \right\} + \frac{1}{\omega_c^4} \sum_{n \neq N, N+2} \frac{n(n-1)}{(N-n)^2(N-n+2)^2}. \tag{S37}$$

Using the relation

$$\sum_{n \neq N, N+2} \frac{n(n-1)}{(N-n)^2(N-n+2)^2} = \frac{2(N^2 + N)\gamma\omega_c^2}{\Gamma^2} - \frac{3}{8}(N^2 + N) + \frac{N}{2} + O(1), \tag{S38}$$

we get

$$\eta_H^I = \frac{\omega_c}{4} \int d\varepsilon (-f_\varepsilon') \nu(\varepsilon) \left[ N^2 + N + N\frac{\varepsilon - \epsilon_N}{\omega_c} - \frac{N^2 + N}{4}\frac{\Gamma^2(\varepsilon)}{\omega_c^2} + N\frac{\Gamma^2}{4\omega_c^2} \right]. \tag{S39}$$

The terms proportional to $N^2$ reproduce the result for $\eta_H^I$ from the main text for well-separated LLs. However, we notice that there is a term of the first order in $N$, $N(\varepsilon - \epsilon_N)/\omega_c$, which could compete with the term $N^2\Gamma^2(\varepsilon)/\omega_c^2$ at moderate $N$. As we shall see below, this term actually cancels with the corresponding term in the expression for $\eta_H^{II}$.

Now we turn to $\eta_H^{II}$. In order to perform such a computation one needs to construct the strain generators in the presence of disorder. We do it perturbatively with respect to the spatial derivatives of a random potential:

$$J_{xy} = \frac{T_{xx} - T_{yy}}{4\omega_c} + \Delta J_{xy}, \qquad J_{xx} - J_{yy} = -\frac{T_{xy}}{\omega_c} + \Delta J_{||}, \tag{S40}$$

where $\Delta J_{xy}$ and $\Delta J_{||}$ satisfy the following equations

$$[H_0, \Delta J_{xy}] = \frac{i}{2\omega_c}\left(\partial_y V v_y - \partial_x V v_x\right) + \frac{1}{4m_e\omega_c}\left(\partial_y^2 V - \partial_x^2 V\right) + \ldots, \tag{S41}$$

$$[H_0, \Delta J_{||}] = -\frac{i}{\omega_c}\left(\partial_x V v_y + \partial_y V v_x\right) - \frac{1}{m_e\omega_c}\partial_y \partial_x V + \ldots. \tag{S42}$$

Then Eq. (S26) can be written as

$$\eta_H^{II} = -\operatorname{Im}\int\frac{d\varepsilon}{8\pi\omega_c S}(-f_\varepsilon')\,\overline{\operatorname{Tr}(T_{xx}-T_{yy})^2 G_\varepsilon^A} - \int\frac{d\varepsilon}{2\pi\omega_c S}f_\varepsilon\overline{\operatorname{Tr} H\operatorname{Im} G_\varepsilon^R}$$
$$+\int\frac{d\varepsilon}{2\pi\omega_c S}f_\varepsilon\overline{\operatorname{Tr} V\operatorname{Im} G_\varepsilon^R} - \operatorname{Im}\int\frac{d\varepsilon}{2\pi S}(-f_\varepsilon')\,\overline{\operatorname{Tr}\Delta J_{xy}(T_{xx}-T_{yy})G_\varepsilon^A}$$
$$+\operatorname{Im}\int\frac{d\varepsilon}{2\pi\omega_c S}f_\varepsilon\overline{\operatorname{Tr}[\Delta J_{xy},T_{xy}]\operatorname{Im} G_\varepsilon^R} - \operatorname{Im}\int\frac{d\varepsilon}{8\pi\omega_c S}f_\varepsilon\overline{\operatorname{Tr}[T_{xx}-T_{yy},\Delta J_{\parallel}]\operatorname{Im} G_\varepsilon^R}. \quad (S43)$$

Here we used the relation (S29). As we shall demonstrate below, the terms on the first line of Eq. (S43) gives contributions of order $N^2$. The terms on the second and third lines of Eq. (S43) are of the order $N$ or less.

Using the matrix elements (S32) we find

$$\operatorname{Im}\int\frac{d\varepsilon}{8\pi\omega_c S}(-f_\varepsilon')\,\overline{\operatorname{Tr}(T_{xx}-T_{yy})^2 G_\varepsilon^A} = \operatorname{Im}\int\frac{d\varepsilon}{8\pi\omega_c S}(-f_\varepsilon')\,\overline{\operatorname{Tr}(T_{xx}-T_{yy})^2\mathcal{G}_\varepsilon^A} = \frac{m_e\omega_c^2}{8\pi^2}\int d\varepsilon f_\varepsilon'\sum_n\frac{(n^2+n+1)\operatorname{Im}\Sigma_\varepsilon^R}{|\varepsilon-\epsilon_n-\Sigma_\varepsilon^R|^2}$$
$$=\frac{(1-\gamma)\omega_c}{4}\int d\varepsilon(-f_\varepsilon')\nu(\varepsilon)\left[N^2+N+\frac{\Gamma^2}{4\omega_c^2}\sum_{n\neq N}{}'\frac{n^2+n+1}{(n-N)^2}\right] = \frac{\omega_c}{4}\int d\varepsilon(-f_\varepsilon')\nu(\varepsilon)\left[N^2+N+\frac{2N\Delta}{\omega_c}\right]. \quad (S44)$$

Here the prime on the sum over $n$ indicates a restriction of the sum to the vicinity of $n=N$, due to the dependence of $\operatorname{Im}\Sigma_\varepsilon^R$ on $\varepsilon-\epsilon_n$.

The second term in Eq. (S43) is estimated as

$$\int\frac{d\varepsilon}{2\pi\omega_c S}f_\varepsilon\overline{\operatorname{Tr} H\operatorname{Im} G_\varepsilon^R} = -\int\frac{d\varepsilon}{2\omega_c}f_\varepsilon\sum_a\overline{E_a\delta(\varepsilon-E_a)} = -\int\frac{d\varepsilon}{2\omega_c}\nu(\varepsilon)\varepsilon f_\varepsilon. \quad (S45)$$

The third term in Eq. (S43) can be approximated as follows:

$$\int\frac{d\varepsilon}{2\pi\omega_c S}f_\varepsilon\overline{\operatorname{Tr} V\operatorname{Im} G_\varepsilon^R} \approx \int\frac{d\varepsilon}{2\pi\omega_c S}f_\varepsilon\operatorname{Im}\operatorname{Tr} V\mathcal{G}_\varepsilon^R V\mathcal{G}_\varepsilon^R = \frac{m_e}{4\pi^2}\int d\varepsilon f_\varepsilon\operatorname{Im}\sum_n\frac{\Sigma_\varepsilon^R}{\varepsilon-\epsilon_n-\Sigma_\varepsilon^R} = \frac{m_e\tau_0}{2\pi\omega_c}\int d\varepsilon f_\varepsilon\operatorname{Im}\left[\Sigma_\varepsilon^R\right]^2$$
$$= -\frac{1}{\omega_c}\int d\varepsilon f_\varepsilon\,\nu(\varepsilon)\operatorname{Re}\Sigma_\varepsilon^R = -\frac{m_e}{\pi^2\Gamma}\int d\varepsilon f_\varepsilon\,\sum_n(\varepsilon-\epsilon_n)\sqrt{1-\frac{(\varepsilon-\epsilon_n)^2}{\Gamma^2}}. \quad (S46)$$

Here the last partially filled LL contributes only in the sum over $n$ at temperatures $T \ll \omega_c$. This implies that this contribution is of order $\Gamma/(l_B^2\omega_c)$, i.e., this term is $O(N^0)$.

Now we consider the fourth term in Eq. (S43):

$$\operatorname{Im}\int\frac{d\varepsilon}{2\pi S}(-f_\varepsilon')\,\overline{\operatorname{Tr}\Delta J_{xy}(T_{xx}-T_{yy})G_\varepsilon^A} \approx \operatorname{Im}\int\frac{d\varepsilon}{2\pi S}(-f_\varepsilon')\,\overline{\operatorname{Tr}\Delta J_{xy}(T_{xx}-T_{yy})\mathcal{G}_\varepsilon^A V\mathcal{G}_\varepsilon^A} = -\frac{m_e\omega_c}{8\pi}\operatorname{Im}\int\frac{d\varepsilon}{2\pi}(-f_\varepsilon')$$
$$\times\sum_{n\neq n'}\frac{1}{n-n'}\sqrt{\frac{n!}{n'!}}\int\frac{d^2\mathbf{Q}}{(2\pi)^2}W(Q)e^{-z_Q}z_Q^{n'-n}L_n^{n'-n}(z_Q)\left[\sqrt{n'(n'-1)}\mathcal{G}_{n'-2}^A(\varepsilon)\mathcal{G}_n^A(\varepsilon)\sqrt{\frac{n!}{(n'-2)!}}L_n^{n'-n-2}(z_Q)\right.$$
$$\left.+\sqrt{(n'+1)(n'+2)}\mathcal{G}_{n'+2}^A(\varepsilon)\mathcal{G}_n^A(\varepsilon)\sqrt{\frac{n!}{(n'+2)!}}z_Q^2 L_n^{n'-n+2}(z_Q)\right] = -\frac{m_e\omega_c}{8\pi}\operatorname{Im}\int\frac{d\varepsilon}{2\pi}(-f_\varepsilon')\sum_{n\neq n'}\frac{1}{n-n'}\frac{n!}{n'!}$$
$$\times\int\frac{d^2\mathbf{Q}}{(2\pi)^2}W(Q)e^{-z_Q}z_Q^{n'-n}L_n^{n'-n}(z_Q)\left\{n'(n'-1)\mathcal{G}_{n'-2}^A(\varepsilon)\mathcal{G}_n^A(\varepsilon)\left[L_n^{n'-n}(z_Q)-2L_{n-1}^{n'-n}(z_Q)+L_{n-2}^{n'-n}(z_Q)\right]\right.$$
$$\left.+\mathcal{G}_{n'+2}^A(\varepsilon)\mathcal{G}_n^A(\varepsilon)\left[(n'+1)(n'+2)(z_Q)-2(n+1)(n'+2)L_{n+1}^{n'-n}(z_Q)+(n+1)(n+2)L_{n+2}^{n'-n}(z_Q)\right]\right\}. \quad (S47)$$

Using asymptotic expression (S8) we obtain

$$
\begin{aligned}
&\operatorname{Im} \int \frac{d\varepsilon}{2\pi S}(-f'_\varepsilon)\, \overline{\operatorname{Tr} \Delta J_{xy}(T_{xx}-T_{yy})G^A_\varepsilon} \approx -\frac{m_e\omega_c}{8\pi}\operatorname{Im}\int \frac{d\varepsilon}{2\pi}(-f'_\varepsilon)\sum_{n\neq n'}\frac{J_{n'-n}(Ql_B\sqrt{n+n'+1})}{n-n'}\int\frac{d^2\boldsymbol{Q}}{(2\pi)^2}W(Q)e^{-z_Q}\\
&\times\Bigg\{\sqrt{n'(n'-1)}\mathcal{G}^A_{n'-2}(\varepsilon)\mathcal{G}^A_n(\varepsilon)\Big[\sqrt{n'(n'-1)}J_{n'-n}(Ql_B\sqrt{n+n'+1})-2\sqrt{n(n'-1)}J_{n'-n}(Ql_B\sqrt{n+n'-1})\\
&+\sqrt{n(n-1)}J_{n'-n}(Ql_B\sqrt{n+n'-3})\Big]+\sqrt{(n'+1)(n'+2)}\mathcal{G}^A_{n'+2}(\varepsilon)\mathcal{G}^A_n(\varepsilon)\Big[\sqrt{(n'+1)(n'+2)}J_{n'-n}(Ql_B\sqrt{n+n'+1})\\
&-2\sqrt{(n+1)(n'+2)}J_{n'-n}(Ql_B\sqrt{n+n'+3})+\sqrt{(n+1)(n+2)}J_{n'-n}(Ql_B\sqrt{n+n'+5})\Big]\Bigg\} = -\frac{m_e\omega_c^2}{16\pi^2}\operatorname{Im}\int\frac{d\varepsilon}{2\pi}(-f'_\varepsilon)\\
&\times\sum_{n\neq n'}\frac{\sqrt{n'(n'-1)}}{n-n'}\Bigg\{\mathcal{G}^A_{n'-2}(\varepsilon)\mathcal{G}^A_n(\varepsilon)\Big[\frac{\sqrt{n'(n'-1)}}{\tau_0}-\frac{2\sqrt{n(n'-1)}}{\tau_1}+\frac{\sqrt{n(n-1)}}{\tau_2}\Big]+\mathcal{G}^A_{n'}(\varepsilon)\mathcal{G}^A_{n-2}(\varepsilon)\Big[\frac{\sqrt{n'(n'-1)}}{\tau_0}\\
&-\frac{2\sqrt{(n-1)n'}}{\tau_1}+\frac{\sqrt{n(n-1)}}{\tau_2}\Big]\Bigg\} \approx \frac{m_e\omega_c^2 N}{8\pi^2\tau_{\text{tr},1}}\operatorname{Im}\int\frac{d\varepsilon}{2\pi}(-f'_\varepsilon)[\mathcal{G}^A_N(\varepsilon)]^2 = \frac{N}{4}\int d\varepsilon(-f'_\varepsilon)\nu(\varepsilon)(\varepsilon-\epsilon_N). \quad (S48)
\end{aligned}
$$

Here we took into account that $\tau_{\text{tr},1}=\tau_0$ for a short-range-correlated random potential.

Next we consider the fifth term in Eq. (S43),

$$
\begin{aligned}
&\operatorname{Im}\int\frac{d\varepsilon}{2\pi\omega_c S}f_\varepsilon\overline{\operatorname{Tr}[\Delta J_{xy},T_{xy}]\operatorname{Im}G^R_\varepsilon} \approx \operatorname{Im}\int\frac{d\varepsilon}{2\pi\omega_c S}f_\varepsilon\overline{\operatorname{Tr}[\Delta J_{xy},T_{xy}](\mathcal{G}^R_\varepsilon V\mathcal{G}^R_\varepsilon-\mathcal{G}^A_\varepsilon V\mathcal{G}^A_\varepsilon)} = \frac{m_e}{2\pi\omega_c S}\operatorname{Im}\int d\varepsilon f_\varepsilon\\
&\times\overline{\operatorname{Tr}[\Delta J_{xy},v_+^2](\mathcal{G}^R_\varepsilon V\mathcal{G}^R_\varepsilon-\mathcal{G}^A_\varepsilon V\mathcal{G}^A_\varepsilon)} = \frac{m_e}{8\pi^2}\operatorname{Im}\int d\varepsilon f_\varepsilon\sum_{n\neq n'}\frac{1}{n-n'}\int\frac{d^2\boldsymbol{Q}}{(2\pi)^2}W(Q)e^{-z_Q}\left(\frac{Q_+l_B}{\sqrt{2}}\right)^{n'-n-2}\\
&\times\Bigg\{\sqrt{n'(n'-1)}\mathcal{G}^R_{n'-2}(\varepsilon)\mathcal{G}^R_n(\varepsilon)\sqrt{\frac{n!}{(n'-2)!}}L_n^{n'-n-2}(z_Q)-\sqrt{(n+1)(n+2)}\mathcal{G}^R_{n+2}(\varepsilon)\mathcal{G}^R_{n'}(\varepsilon)\sqrt{\frac{(n+2)!}{n'!}}L_{n+2}^{n'-n-2}(z_Q)\Bigg\}\\
&\times\Bigg\{\sqrt{\frac{n!}{n'!}}\frac{Q_+l_B}{\sqrt{2}}\left(\frac{Q_-l_B}{\sqrt{2}}\right)^{n'-n+1}L_n^{n'-n+1}(z_Q)+\sqrt{\frac{n!n'}{(n'-1)!}}\frac{Q_-l_B}{\sqrt{2}}\left(\frac{Q_-l_B}{\sqrt{2}}\right)^{n'-n-1}L_n^{n'-n-1}(z_Q)\\
&-\frac{(Q_+^2+Q_-^2)l_B^2}{4}\sqrt{\frac{n!}{n'!}}\left(\frac{Q_-l_B}{\sqrt{2}}\right)^{n'-n}L_n^{n'-n}(z_Q)\Bigg\} = -\frac{m_e}{16\pi^2}\operatorname{Im}\int d\varepsilon f_\varepsilon\sum_{n\neq n'}\frac{1}{n-n'}\frac{n!}{n'!}\int\frac{d^2\boldsymbol{Q}}{(2\pi)^2}W(Q)e^{-z_Q}z_Q^{n'-n}\\
&\times L_n^{n'-n}(z_Q)\Bigg\{n'(n'-1)\mathcal{G}^R_{n'-2}(\varepsilon)\mathcal{G}^R_n(\varepsilon)\Big[L_n^{n'-n}(z_Q)-2L_{n-1}^{n'-n}(z_Q)+L_{n-2}^{n'-n}(z_Q)\Big]-(n+1)(n+2)\mathcal{G}^R_{n'}(\varepsilon)\mathcal{G}^R_{n+2}(\varepsilon)\\
&\times\Big[L_n^{n'-n}(z_Q)-2L_{n+1}^{n'-n}(z_Q)+L_{n+2}^{n'-n}(z_Q)\Big]\Bigg\} = -\frac{m_e}{16\pi^2}\operatorname{Im}\int d\varepsilon f_\varepsilon\sum_{n\neq n'}\frac{\mathcal{G}^R_{n'-2}\mathcal{G}^R_n}{n-n'}\int\frac{d^2\boldsymbol{Q}}{(2\pi)^2}W(Q)\\
&\times\Big[\sqrt{n'(n'-1)}J_{n'-n}(Ql_B\sqrt{n+n'+1})-\sqrt{n(n-1)}J_{n'-n}(Ql_B\sqrt{n+n'-3})\Big]\Big[\sqrt{n'(n'-1)}J_{n'-n}(Ql_B\sqrt{n+n'+1})\\
&-2\sqrt{n(n'-1)}J_{n'-n}(Ql_B\sqrt{n+n'-1})+\sqrt{n(n-1)}J_{n'-n}(Ql_B\sqrt{n+n'-3})\Big] = \frac{m_e\omega_c}{16\pi^2}\operatorname{Im}\int\frac{d\varepsilon}{2\pi}f_\varepsilon\sum_{n\neq n'}(n'+n-1)\\
&\times\mathcal{G}^R_{n'-2}\mathcal{G}^R_n\left(\frac{1}{\tau_0}-\frac{1}{\tau_1}\frac{2\sqrt{n(n'-1)}}{\sqrt{n(n-1)}+\sqrt{n'(n'-1)}}\right) \approx \frac{m_e\omega_c}{16\pi^2\tau_0}\operatorname{Im}\int\frac{d\varepsilon}{2\pi}f_\varepsilon\sum_n(2n+1)\left(\mathcal{G}^R_n(\varepsilon)\right)^2\\
&= \frac{m_e}{8\pi^2\Gamma}\int d\varepsilon f_\varepsilon\,(\varepsilon-\epsilon_n)\sqrt{1-\frac{(\varepsilon-\epsilon_n)^2}{\Gamma^2}}. \quad (S49)
\end{aligned}
$$

Here we took into account that $1/\tau_1=0$ for the case of the short-range-correlated random potential. For temperature $T\ll\omega_c$ the last partially filled Landau level contributes only to the sum over $n$. This implies that this contribution is of the order of $N\Gamma/(l_B^2\omega_c)$, i.e., can be safely neglected.

Finally, the last term in Eq. (S43) can be estimated as

$$\operatorname{Im} \int \frac{d\varepsilon}{8\pi\omega_c S} f_\varepsilon \overline{\operatorname{Tr}[T_{xx} - T_{yy}, \Delta(J_{xx} - J_{yy})] \operatorname{Im} G_\varepsilon^R} = \frac{m_e}{4\pi\omega_c S} \operatorname{Re} \int d\varepsilon f_\varepsilon \overline{\operatorname{Tr}[\Delta(J_{xx} - J_{yy}), v_+^2](\mathcal{G}_\varepsilon^R V \mathcal{G}_\varepsilon^R - \mathcal{G}_\varepsilon^A V \mathcal{G}_\varepsilon^A)}$$

$$= -\operatorname{Re} \frac{m_e}{16\pi^2 i} \int d\varepsilon f_\varepsilon \sum_{n \neq n'} \frac{1}{n-n'} \int \frac{d^2\mathbf{Q}}{(2\pi)^2} W(Q) e^{-z_Q} z_Q^{n'-n} \frac{n!}{n'!} L_n^{n'-n}(z_Q) \left\{ n'(n'-1)\mathcal{G}_{n'-2}^R(\varepsilon)\mathcal{G}_n^R(\varepsilon) \left[L_n^{n'-n}(z_Q) \right. \right.$$

$$\left. \left. -2L_{n-1}^{n'-n}(z_Q) + L_{n-2}^{n'-n}(z_Q)\right] - (n+1)(n+2)\mathcal{G}_{n'}^R(\varepsilon)\mathcal{G}_{n+2}^R(\varepsilon) \left[L_n^{n'-n}(z_Q) - 2L_{n+1}^{n'-n}(z_Q) + L_{n+2}^{n'-n}(z_Q)\right] \right\}$$

$$= \frac{m_e \omega_c}{16\pi^2} \operatorname{Im} \int \frac{d\varepsilon}{2\pi} f_\varepsilon \sum_{n \neq n'} (n' + n - 1)\mathcal{G}_{n'-2}^R(\varepsilon)\mathcal{G}_n^R(\varepsilon) \left(\frac{1}{\tau_0} - \frac{1}{\tau_1} \frac{2\sqrt{n(n'-1)}}{\sqrt{n(n-1)} + \sqrt{n'(n'-1)}}\right)$$

$$\approx \frac{m_e \omega_c}{16\pi^2 \tau_0} \operatorname{Im} \int \frac{d\varepsilon}{2\pi} f_\varepsilon \sum_n (2n+1) \left(\mathcal{G}_n^R(\varepsilon)\right)^2 = \frac{m_e}{8\pi^2 \Gamma} \int d\varepsilon f_\varepsilon (\varepsilon - \epsilon_n) \sqrt{1 - \frac{(\varepsilon - \epsilon_n)^2}{\Gamma^2}}. \quad (S50)$$

Therefore, the last term is the same as penultimate one.

Combining together the results (S44)-(S50), we find

$$\eta_H^{II} = \frac{1}{2\omega_c} \int d\varepsilon \nu(\varepsilon) \varepsilon f_\varepsilon - \frac{\omega_c}{4} \int d\varepsilon (-f_\varepsilon') \nu(\varepsilon) \left[N^2 + N \frac{\varepsilon - \epsilon_N}{\omega_c}\right]. \quad (S51)$$

Finally, using Eq. (S39) and (S51), we get

$$\eta_H = \frac{1}{2\omega_c} \int d\varepsilon \nu(\varepsilon) \varepsilon f_\varepsilon - \frac{N^2}{16\omega_c} \int d\varepsilon (-f_\varepsilon') \nu(\varepsilon) \Gamma^2(\varepsilon). \quad (S52)$$

## S.VI. NUMERICAL METHOD

In the numerical calculation, we employ the Hofstadter model on a square lattice of size $N_x \times N_y$ with periodic boundary conditions in the presence of disorder:

$$\hat{H} = \sum_{n_x=1}^{N_x} \sum_{n_y=1}^{N_y} V_{n_x,n_y} |n_x, n_y\rangle \langle n_x, n_y| - t |n_x + 1, n_y\rangle \langle n_x, n_y| - t e^{-i2\pi\alpha n_x} |n_x, n_y + 1\rangle \langle n_x, n_y| + \text{H.c.}, \quad (S53)$$

where $|n_x, n_y\rangle$ denotes the orbital on site $(n_x, n_y)$, $V_{n_x,n_y}$ are independent random potential amplitudes, uniformly distributed between $[-w/2, w/2]$, $t = \hbar^2/2m_e a^2$ is the nearest-neighbor hopping amplitude, and $\alpha = eBa^2/2\pi\hbar c$ is the flux per plaquette in units of the flux quantum, $a$ being the lattice spacing. We employ a lattice expression for the stress tensor operator, which reduces to the usual expression in the continuum limit [S9]:

$$\hat{T}_{jk} = -t \left(\hat{D}_i \hat{D}_j + \hat{D}_j \hat{D}_i\right), \quad (S54)$$

where,

$$\hat{D}_x = \sum_{n_x,n_y} |n_x + 1, n_y\rangle \langle n_x, n_y| - |n_x - 1, n_y\rangle \langle n_x, n_y|, \quad (S55)$$

$$\hat{D}_y = \sum_{n_x,n_y} |n_x, n_y\rangle \langle n_x, n_y + 1| e^{-i2\pi\alpha n_x} - |n_x, n_y - 1\rangle \langle n_x, n_y| e^{i2\pi\alpha n_x}. \quad (S56)$$

Denoting the single-particle eigenenergies and eigenstates of the Hofstadter Hamiltonian by $E_\mu$ and $|\mu\rangle$, respectively, the Hall viscosity at zero temperature is then calculated from the Kubo formula as a sum over the occupied and unoccupied eigenstates:

$$\frac{\hbar}{m_e} \eta_H = -2t \operatorname{Im} \sum_{\substack{E_{\mu_1} < E_F, \\ E_{\mu_2} > E_F}} \frac{\langle \mu_2 | \hat{T}_{xx} | \mu_1 \rangle \langle \mu_1 | \hat{T}_{xy} | \mu_2 \rangle}{(E_{\mu_1} - E_{\mu_2})^2}. \quad (S57)$$




\begin{thebibliography}{100}
\bibitem{note:rotation}
\color{black}
Up to small mesoscopic fluctuations (see, e.g., Ref.~\cite{anisovich87} for the case of the thermopower), as we have also verified numerically.
\end{thebibliography}
\end{document}